\pdfoutput=1 

\documentclass[twocolumn,aps,prc,superscriptaddress,nofootinbib,showpacs,floatfix]{revtex4}

\usepackage{graphicx}	
\usepackage{xspace}     

\newcommand{\pt}{\mbox{$p_T$}\xspace}

\newcommand{\sqsn}{\mbox{$\sqrt{s_{_{NN}}}$}\xspace}
\newcommand{\ee}  {\mbox{$e^+e^-$}\xspace}
\newcommand{\dA}  {\mbox{$d$$+$Au}\xspace}
\newcommand{\pp}  {\mbox{$p$$+$$p$}\xspace}

\newcommand{\pythia}{{\sc pythia}}
\newcommand{\mcnlo}{{\sc mc@nlo}}
\newcommand{\herwig}{{\sc Herwig}}
\newcommand{\geant}{{\sc geant}3}
\newcommand{\cc}{\mbox{$c\bar{c}$}\xspace}
\newcommand{\bb}{\mbox{$b\bar{b}$}\xspace}
\newcommand{\qq}{\mbox{$q\bar{q}$}\xspace}
\newcommand{\Nunlike}{\mbox{$N_{+-}$}}
\newcommand{\Nlike}{\mbox{$N_{\pm\pm}$}}
\newcommand{\Blike}{\mbox{$B^{\rm comb}_{\pm\pm}$}}
\newcommand{\Bunlike}{\mbox{$B^{\rm comb}_{+-}$}}

\newcommand{\Bcorr}{\mbox{$B^{\rm cor}_{+-}$}}
\newcommand{\signal}{\mbox{$S_{+-}$}}

\begin{document}


\title{Cross section for $b\bar{b}$ production via dielectrons in 
\dA collisions at $\sqrt{s_{_{NN}}}=$200~GeV}

\newcommand{\abilene}{Abilene Christian University, Abilene, Texas 79699, USA}
\newcommand{\augie}{Department of Physics, Augustana College, Sioux Falls, South Dakota 57197, USA}
\newcommand{\banaras}{Department of Physics, Banaras Hindu University, Varanasi 221005, India}
\newcommand{\barc}{Bhabha Atomic Research Centre, Bombay 400 085, India}
\newcommand{\baruch}{Baruch College, City University of New York, New York, New York, 10010 USA}
\newcommand{\bnlcoll}{Collider-Accelerator Department, Brookhaven National Laboratory, Upton, New York 11973-5000, USA}
\newcommand{\bnlphys}{Physics Department, Brookhaven National Laboratory, Upton, New York 11973-5000, USA}
\newcommand{\caucr}{University of California - Riverside, Riverside, California 92521, USA}
\newcommand{\charlesczech}{Charles University, Ovocn\'{y} trh 5, Praha 1, 116 36, Prague, Czech Republic}
\newcommand{\chonbuk}{Chonbuk National University, Jeonju, 561-756, Korea}
\newcommand{\ciae}{Science and Technology on Nuclear Data Laboratory, China Institute of Atomic Energy, Beijing 102413, P.~R.~China}
\newcommand{\cns}{Center for Nuclear Study, Graduate School of Science, University of Tokyo, 7-3-1 Hongo, Bunkyo, Tokyo 113-0033, Japan}
\newcommand{\colorado}{University of Colorado, Boulder, Colorado 80309, USA}
\newcommand{\columbia}{Columbia University, New York, New York 10027 and Nevis Laboratories, Irvington, New York 10533, USA}
\newcommand{\czechtech}{Czech Technical University, Zikova 4, 166 36 Prague 6, Czech Republic}
\newcommand{\dapnia}{Dapnia, CEA Saclay, F-91191, Gif-sur-Yvette, France}
\newcommand{\elte}{ELTE, E{\"o}tv{\"o}s Lor{\'a}nd University, H - 1117 Budapest, P{\'a}zm{\'a}ny P. s. 1/A, Hungary}
\newcommand{\ewha}{Ewha Womans University, Seoul 120-750, Korea}
\newcommand{\fit}{Florida Institute of Technology, Melbourne, Florida 32901, USA}
\newcommand{\fsu}{Florida State University, Tallahassee, Florida 32306, USA}
\newcommand{\gsu}{Georgia State University, Atlanta, Georgia 30303, USA}
\newcommand{\hiroshima}{Hiroshima University, Kagamiyama, Higashi-Hiroshima 739-8526, Japan}
\newcommand{\ihepprot}{IHEP Protvino, State Research Center of Russian Federation, Institute for High Energy Physics, Protvino, 142281, Russia}
\newcommand{\illuiuc}{University of Illinois at Urbana-Champaign, Urbana, Illinois 61801, USA}
\newcommand{\inrras}{Institute for Nuclear Research of the Russian Academy of Sciences, prospekt 60-letiya Oktyabrya 7a, Moscow 117312, Russia}
\newcommand{\instpasczech}{Institute of Physics, Academy of Sciences of the Czech Republic, Na Slovance 2, 182 21 Prague 8, Czech Republic}
\newcommand{\isu}{Iowa State University, Ames, Iowa 50011, USA}
\newcommand{\jaea}{Advanced Science Research Center, Japan Atomic Energy Agency, 2-4 Shirakata Shirane, Tokai-mura, Naka-gun, Ibaraki-ken 319-1195, Japan}
\newcommand{\jyvaskyla}{Helsinki Institute of Physics and University of Jyv{\"a}skyl{\"a}, P.O.Box 35, FI-40014 Jyv{\"a}skyl{\"a}, Finland}
\newcommand{\kek}{KEK, High Energy Accelerator Research Organization, Tsukuba, Ibaraki 305-0801, Japan}
\newcommand{\korea}{Korea University, Seoul, 136-701, Korea}
\newcommand{\kurchatov}{Russian Research Center ``Kurchatov Institute", Moscow, 123098 Russia}
\newcommand{\kyoto}{Kyoto University, Kyoto 606-8502, Japan}
\newcommand{\labllr}{Laboratoire Leprince-Ringuet, Ecole Polytechnique, CNRS-IN2P3, Route de Saclay, F-91128, Palaiseau, France}
\newcommand{\lahorelums}{Physics Department, Lahore University of Management Sciences, Lahore 54792, Pakistan}
\newcommand{\lawllnl}{Lawrence Livermore National Laboratory, Livermore, California 94550, USA}
\newcommand{\losalamos}{Los Alamos National Laboratory, Los Alamos, New Mexico 87545, USA}
\newcommand{\lpc}{LPC, Universit{\'e} Blaise Pascal, CNRS-IN2P3, Clermont-Fd, 63177 Aubiere Cedex, France}
\newcommand{\lund}{Department of Physics, Lund University, Box 118, SE-221 00 Lund, Sweden}
\newcommand{\maryland}{University of Maryland, College Park, Maryland 20742, USA}
\newcommand{\mass}{Department of Physics, University of Massachusetts, Amherst, Massachusetts 01003-9337, USA }
\newcommand{\michigan}{Department of Physics, University of Michigan, Ann Arbor, Michigan 48109-1040, USA}
\newcommand{\muenster}{Institut fur Kernphysik, University of Muenster, D-48149 Muenster, Germany}
\newcommand{\muhlenberg}{Muhlenberg College, Allentown, Pennsylvania 18104-5586, USA}
\newcommand{\myongji}{Myongji University, Yongin, Kyonggido 449-728, Korea}
\newcommand{\nagasaki}{Nagasaki Institute of Applied Science, Nagasaki-shi, Nagasaki 851-0193, Japan}
\newcommand{\newmex}{University of New Mexico, Albuquerque, New Mexico 87131, USA }
\newcommand{\nmsu}{New Mexico State University, Las Cruces, New Mexico 88003, USA}
\newcommand{\ohio}{Department of Physics and Astronomy, Ohio University, Athens, Ohio 45701, USA}
\newcommand{\ornl}{Oak Ridge National Laboratory, Oak Ridge, Tennessee 37831, USA}
\newcommand{\orsay}{IPN-Orsay, Universite Paris Sud, CNRS-IN2P3, BP1, F-91406, Orsay, France}
\newcommand{\peking}{Peking University, Beijing 100871, P.~R.~China}
\newcommand{\pnpi}{PNPI, Petersburg Nuclear Physics Institute, Gatchina, Leningrad region, 188300, Russia}
\newcommand{\riken}{RIKEN Nishina Center for Accelerator-Based Science, Wako, Saitama 351-0198, Japan}
\newcommand{\rikjrbrc}{RIKEN BNL Research Center, Brookhaven National Laboratory, Upton, New York 11973-5000, USA}
\newcommand{\rikkyo}{Physics Department, Rikkyo University, 3-34-1 Nishi-Ikebukuro, Toshima, Tokyo 171-8501, Japan}
\newcommand{\saopaulo}{Universidade de S{\~a}o Paulo, Instituto de F\'{\i}sica, Caixa Postal 66318, S{\~a}o Paulo CEP05315-970, Brazil}
\newcommand{\stonybrkc}{Chemistry Department, Stony Brook University, SUNY, Stony Brook, New York 11794-3400, USA}
\newcommand{\stonycrkp}{Department of Physics and Astronomy, Stony Brook University, SUNY, Stony Brook, New York 11794-3800,, USA}
\newcommand{\tenn}{University of Tennessee, Knoxville, Tennessee 37996, USA}
\newcommand{\titech}{Department of Physics, Tokyo Institute of Technology, Oh-okayama, Meguro, Tokyo 152-8551, Japan}
\newcommand{\tsukuba}{Institute of Physics, University of Tsukuba, Tsukuba, Ibaraki 305, Japan}
\newcommand{\vandy}{Vanderbilt University, Nashville, Tennessee 37235, USA}
\newcommand{\waseda}{Waseda University, Advanced Research Institute for Science and Engineering, 17 Kikui-cho, Shinjuku-ku, Tokyo 162-0044, Japan}
\newcommand{\weizmann}{Weizmann Institute, Rehovot 76100, Israel}
\newcommand{\wigner}{Institute for Particle and Nuclear Physics, Wigner Research Centre for Physics, Hungarian Academy of Sciences (Wigner RCP, RMKI) H-1525 Budapest 114, POBox 49, Budapest, Hungary}
\newcommand{\yonsei}{Yonsei University, IPAP, Seoul 120-749, Korea}
\affiliation{\abilene}
\affiliation{\augie}
\affiliation{\banaras}
\affiliation{\barc}
\affiliation{\baruch}
\affiliation{\bnlcoll}
\affiliation{\bnlphys}
\affiliation{\caucr}
\affiliation{\charlesczech}
\affiliation{\chonbuk}
\affiliation{\ciae}
\affiliation{\cns}
\affiliation{\colorado}
\affiliation{\columbia}
\affiliation{\czechtech}
\affiliation{\dapnia}
\affiliation{\elte}
\affiliation{\ewha}
\affiliation{\fit}
\affiliation{\fsu}
\affiliation{\gsu}
\affiliation{\hiroshima}
\affiliation{\ihepprot}
\affiliation{\illuiuc}
\affiliation{\inrras}
\affiliation{\instpasczech}
\affiliation{\isu}
\affiliation{\jaea}
\affiliation{\jyvaskyla}
\affiliation{\kek}
\affiliation{\korea}
\affiliation{\kurchatov}
\affiliation{\kyoto}
\affiliation{\labllr}
\affiliation{\lahorelums}
\affiliation{\lawllnl}
\affiliation{\losalamos}
\affiliation{\lpc}
\affiliation{\lund}
\affiliation{\maryland}
\affiliation{\mass}
\affiliation{\michigan}
\affiliation{\muenster}
\affiliation{\muhlenberg}
\affiliation{\myongji}
\affiliation{\nagasaki}
\affiliation{\newmex}
\affiliation{\nmsu}
\affiliation{\ohio}
\affiliation{\ornl}
\affiliation{\orsay}
\affiliation{\peking}
\affiliation{\pnpi}
\affiliation{\riken}
\affiliation{\rikjrbrc}
\affiliation{\rikkyo}
\affiliation{\saopaulo}
\affiliation{\stonybrkc}
\affiliation{\stonycrkp}
\affiliation{\tenn}
\affiliation{\titech}
\affiliation{\tsukuba}
\affiliation{\vandy}
\affiliation{\waseda}
\affiliation{\weizmann}
\affiliation{\wigner}
\affiliation{\yonsei}
\author{A.~Adare} \affiliation{\colorado}
\author{C.~Aidala} \affiliation{\mass} \affiliation{\michigan}
\author{N.N.~Ajitanand} \affiliation{\stonybrkc}
\author{Y.~Akiba} \affiliation{\riken} \affiliation{\rikjrbrc}
\author{H.~Al-Bataineh} \affiliation{\nmsu}
\author{J.~Alexander} \affiliation{\stonybrkc}
\author{A.~Angerami} \affiliation{\columbia}
\author{K.~Aoki} \affiliation{\kyoto} \affiliation{\riken}
\author{N.~Apadula} \affiliation{\stonycrkp}
\author{Y.~Aramaki} \affiliation{\cns} \affiliation{\riken}
% ...
\collaboration{PHENIX Collaboration} \noaffiliation

\date{\today}

\begin{abstract}


We report a measurement of $e^+e^-$ pairs from semileptonic 
heavy-flavor decays in $d$$+$Au collisions at $\sqrt{s_{_{NN}}}=200$~GeV. 
Exploring the mass and transverse-momentum dependence of the yield, 
the bottom decay contribution can be isolated from charm, and 
quantified by comparison to {\sc pythia} and {\sc mc@nlo} 
simulations.  The resulting $b\bar{b}$-production cross section is
$\sigma^{d{\rm Au}}_{b\bar{b}}=1.37{\pm}0.28({\rm stat}){\pm}0.46({\rm syst})$~mb, 
which is equivalent to a nucleon-nucleon cross section of 
$\sigma^{NN}_{bb}=3.4\pm0.8({\rm stat}){\pm}1.1({\rm syst})\ \mu$b.

\end{abstract}

	
\maketitle

\section{Introduction}

Collisions of heavy nuclei at the Relativistic Heavy Ion Collider 
(RHIC) at Brookhaven National Laboratory produce a quark-gluon 
plasma,  which is a fundamentally new strongly coupled 
state of partonic matter~\cite{whitepaper, starwp, phoboswp, 
brahmswp}. There is extensive experimental evidence that partons lose 
energy while traversing the hot medium~\cite{ppg003, ppg014, 
starhipt}. Many theoretical studies have been performed to determine 
the role of gluon radiation and collisional energy loss 
processes~\cite{charmeloss, charmeloss2}, as well as to confront the 
data with predictions based upon AdS/CFT~\cite{adscft}. 

The fate of a higher mass quark traversing the plasma can help 
elucidate the mechanism of the energy loss, as the quark mass affects 
gluon radiation in the medium~\cite{dima_deadcone}. Consequently, 
single electrons and positrons from the decays of mesons containing 
heavy quarks have been studied in various systems at both RHIC 
\cite{ppg077, ppg131, STARb} and the Large Hadron Collider at CERN 
\cite{alice1, alice2}.

Differentiating among theoretical descriptions of the energy loss 
will be aided by comparing charm and bottom yields. In order to 
observe quark-gluon plasma effects on heavy quarks, it is crucial to 
compare Au$+$Au data to a baseline measurement not dominated by the 
plasma. Typically \pp collisions are used to provide this baseline. 
There are also effects of cold nuclear matter on the production of 
heavy quarks, which can be studied by comparing \pp to $p$$+$Pb or 
$d$$+$Au. PHENIX has already reported modification in cold nuclear 
matter of single electrons at moderate $p_T$ \cite{ppg131}, heavy 
flavor measured through $e$-$\mu$ correlations \cite{PPG130} and 
$J/\psi$ \cite{ppg125, ppg151}. Of course, the bound state can be 
broken up in cold nuclear matter, so the \cc and \bb production cross 
sections in \dA are of interest.

Clean c/b separation is difficult to achieve with single lepton 
measurements, as the single lepton spectrum contains both charm and 
bottom contributions. The B decay contribution increases with $p_T$, 
and is comparable to the D decay contribution at $p_T \ge$ 3 GeV/$c$ 
\cite{ppg094,STAReh}. PHENIX performed initial measurements of the 
charm and bottom cross sections in \pp collisions via high mass 
dielectrons \cite{ppg085} and electron-hadron correlations 
\cite{ppg094}. STAR also reported a \bb cross section in \pp 
collisions \cite{STARb} measured through single electron spectra.

Reconstructing heavy flavor hadrons or measuring leptons with 
displaced vertices allows more direct separation of charm and bottom. 
However, such measurements require microvertex detectors or large 
data sets into a very large aperture with high resolution hadron 
identification. PHENIX has a new silicon microvertex detector, but no 
$d+$Au data have been collected with it yet.

Dielectron spectra, which are double differential in mass and $p_T$, 
allow separation of regions dominated by charm from those dominated 
by bottom. The yield and shape of the mass and $p_T$ spectra provide 
sensitivity to the heavy flavor cross sections. Furthermore, the 
spectra can also encode information about the heavy flavor production 
mechanism via the dielectron correlations, which affect the detected 
pair mass and $p_T$ and therefore the spectral shape.

Initial-state effects such as gluon shadowing in the nucleus may 
affect heavy quark cross sections as the dominant production channel 
at RHIC is gluon fusion. The shape of the mass and $p_T$ 
distributions of charm and bottom decay electrons could additionally 
be sensitive to other effects, such as parton energy loss and 
rescattering in cold nuclear matter, for which evidence was recently 
reported \cite{ppg131}. While azimuthal correlations of the two 
leptons have advantages for studying the heavy-quark production 
process~\cite{PPG130}, analysis of dileptons as a function of mass 
and \pt is undertaken in order to separate charm and bottom 
contributions.

In this paper we report a high statistics measurement of dielectrons 
in \dA collisions in order to provide part of the necessary baseline 
information for quark-gluon plasma studies. Section II describes the 
experimental apparatus and trigger. Section III presents details 
about the data analysis, including electron identification, 
background subtraction, and efficiency corrections. The data are 
presented in Section IV, as double differential spectra in mass and 
$p_T$. Expected sources of dielectrons, and effects of the PHENIX 
acceptance are also discussed in this section. In Section V the 
results are compared to models of charm and bottom production to 
determine the heavy flavor cross sections and examine sensitivity to 
leading-order and next-to-leading-order quantum chromodynamics (QCD) 
descriptions of heavy-flavor physics. Section VI presents our summary 
and conclusions.

\section{Experiment}

The data reported in this paper were collected in the 2008 RHIC \dA 
run. The data were recorded by the PHENIX detector using both a 
minimum bias trigger and an electron trigger. A total of 3.1 billion 
triggered events were analyzed, corresponding to 116.6 billion 
sampled minimum bias events and an integrated luminosity of 
$58.6~n$b$^{-1}$ (equivalent to a nucleon-nucleon $\int \! L dt=23 
~p$b$^{-1}$).

Electrons are measured in PHENIX using the two central arm 
spectrometers, each covering $|\eta| < $ 0.35 and $\Delta\phi = 
\pi/2$. A detailed description of the PHENIX detector is available 
in~\cite{mainNIM}. Tracks are reconstructed using information from 
hits in the drift chambers (DC) and pad chambers. The magnitude 
of the particle's bend in the central axial magnetic field is 
determined from the reconstructed track and used to determine the 
track's momentum. The momentum resolution for this data set is 
$\delta p/p$ = 0.011 $\oplus$ 0.0116$p$, where p is in GeV/$c$.

Tracks are projected onto the photomultiplier tube plane of the 
ring-imaging \v{C}erenkov counter (RICH). Matched hits allow cuts on 
the ring shape and size to separate electrons from hadrons to 
approximately 5 GeV/$c$. The electromagnetic calorimeters (EMCal) 
measure the deposited energy and the shower shape. The ratio of the 
measured energy and momentum provides further electron 
identification~\cite{ppg077}.

The collision vertex, collision time, and minimum bias trigger are 
provided by a pair of beam-beam counters (BBC) located 144 cm from 
the center of PHENIX, on either side of the collision region. Each 
BBC comprises 64 quartz \v{C}erenkov counters and covers a rapidity 
range of 3.0 $< |\eta| <$ 3.9. The collision vertex resolution is 
approximately 0.5 cm in \dA collisions. The minimum bias trigger 
requires a coincidence between North and South sides of the BBC, with 
at least one hit on each side and accepts the events if the BBC 
vertex is within 38 $cm$ of the nominal interaction vertex. The 
minimum bias trigger is sensitive to $88\pm 4$\% of all \dA 
collisions \cite{ppg160}.

Collisions producing an electron-positron pair are extremely rare; 
fewer than 1\% of minimum bias triggered events contain a single 
electron (\pt $>$200 MeV) in the central arm acceptance. 
Consequently, pairs exist in only a tiny fraction of the events. 
Furthermore, pairs at high mass and high $p_T$ have cross sections 
many orders of magnitude smaller than pairs from vector meson decays. 
As a result, electron triggered events are crucial for collecting a 
high statistics dielectron sample in \dA. The PHENIX electron trigger 
requires a \v{C}erenkov ring deposited in the RICH that is spatially 
aligned with a shower in the EMCal with energy above two thresholds 
of 600 and 800 MeV. The bias of the resulting samples is corrected 
for the trigger efficiency using the ratio of electron triggered to 
minimum bias triggered events as a function of electron momentum. The 
minimum bias triggered data sample does not require any minimum 
energy threshold criterion. The single electron trigger threshold 
creates a mass threshold for electron pairs, and is corrected by 
comparing to minimum bias collisions double-differentially in mass 
and $p_T$.

\section{Data Analysis}

Data quality cuts include fiducial cuts to remove any detector edge 
effects or dead areas. The data were collected into run groups with 
similar detector performance characteristics. Each group was analyzed 
separately, and the groups were combined after efficiency correction.

\subsection{Electron identification}

Electron candidates must pass track reconstruction quality cuts, have 
$p_T >$ 0.2 GeV/$c$, as well as fire the RICH and EMCal detectors. To 
be identified as an electron, each candidate must be associated with 
two or more fired RICH photomultiplier tubes within the expected RICH 
ring size and position. In the relatively low multiplicity $d + $Au 
collisions, this is the main discriminating cut using the RICH. 
Electrons are also required to have a good match to an EMCal cluster. 
For further electron identification, the energy in the EMCal must 
satisfy the requirement $E/p >$ 0.5. The electron purity is 
approximately 85-90\%. Finally, to fully control the kinematic edge 
of the single electron $p_{T}$ cut, the pair 
$m_{T}=\sqrt{m^{2}+p_{T}^{2}}$ is required to be greater 
than 450~MeV/$c$.

Photon conversions in detector support structures are identified in 
the two-dimensional plane of DC hit azimuthal angle vs. $E/p$. 
Conversion electrons traverse a portion of the magnetic field and, 
consequently, their momentum and, therefore, their $E/p$ is 
mismeasured. Fully reconstructed conversions in the beam pipe and air 
before the DC are removed by a cut on a pairwise variable, $\phi_V$, 
defined as

\begin{eqnarray}
  \label{eq:phivdef}
  \vec{u}&=&\frac{\vec{p}_1 + \vec{p}_2}{|\vec{p}_1 + \vec{p}_2|},\\
  \vec{v}&=&\vec{p}_1 \times \vec{p}_2,\\
  \vec{w}&=&\vec{u} \times \vec{v},\\
\vec{u}_a&=&\frac{\vec{u} \times \hat{z}}{|\vec{u} \times \hat{z}|},\\
  \phi_V&=&\arccos \left(\frac{\vec{w} \cdot \vec{u}_a}{|\vec{w}| |\vec{u}_a|}\right).
 \end{eqnarray}

\noindent Here $\vec{p}_1$ is the 3-momentum vector of the electron 
and $\vec{p}_2$ the 3-momentum vector of the positron. This is a cut 
on the orientation of the plane defined by the opening angle of the 
pair with respect to the magnetic field, which is parallel to the 
beam axis $\vec{z}$.  The $e^{+}e^{-}$ pairs from photon conversions 
have no intrinsic opening angle. Therefore, the only way they can be 
separated from each other is by the magnetic field pulling them 
apart.  In this case, the opening angle will be aligned perpendicular 
to the magnetic field.  However, any pair that decays from a source 
with mass must have an opening angle that is randomly oriented with 
respect to the magnetic field. For $m_{ee}<$600 MeV/c$^2$, this cut 
removes 98\% of the conversions while retaining 80\% of the signal 
pairs. At higher pair mass where the heavy flavor spectrum dominates, 
conversions are negligible and this cut does not affect the signal 
efficiency.

An additional source of contamination in the dielectron spectrum is 
due to hadron tracks that share a RICH ring with an electron. The 
sharing cannot be properly reproduced by event-mixing, so this 
contamination must be removed before background subtraction. As 
like-sign electron-hadron pairs populate a different region in mass 
and $p_T$ from unlike-sign pairs, like-sign subtraction also cannot 
be used to remove this contamination. Consequently, a cut is placed 
on the distance between the projection of any two tracks onto the 
RICH photomultiplier tube plane. If the projections are within 10 
$\sigma$ in $\Delta\phi_{\rm RICH} \oplus \Delta z_{\rm RICH}$ (this 
corresponds $\approx$ 36 $cm$, roughly twice the predicted maximum 
diameter of a RICH ring), then the entire event is rejected. This cut 
does not affect the mass spectrum above $m_{ee} >600$~MeV/$c^2$ and 
removes less than 1\% of the events.

\subsection{Background Subtraction}

All electrons and positrons in a given event are combined into pairs. 
We refer to these as {\it foreground} and denote the number of \ee 
pairs as \Nunlike\ and the like-sign pairs as \Nlike. The foreground 
pairs contain signal pairs (\signal) from the sources that we are 
interested in, and background pairs.  Electrons and positrons from 
different physical sources (\Bunlike) are uncorrelated. Additionally, 
there are some \ee background pairs which are correlated (\Bcorr), 
described in Section \ref{sec:corr_bkg}. Both types of background are 
subtracted statistically from the foreground to extract the signal.

Since the background is typically larger than the signal, the 
background estimation requires precision of a few percent. The 
signal-to-background ($S/B$) ratio varies with invariant mass of the 
pairs. In \dA collisions, the $p_{T}$ integrated $S/B$ is larger than 
1.0 only near the vector meson masses. It is below 0.1 for the low 
mass continuum ($<$1.0 GeV/$c^2$). In the intermediate mass continuum 
(1.0-3.0 GeV/$c^2$), the $S/B$ is roughly constant between 0.2-0.3; 
the $S/B$ increases for higher mass.

\begin{figure}[htb]
\includegraphics[width=0.95\linewidth]{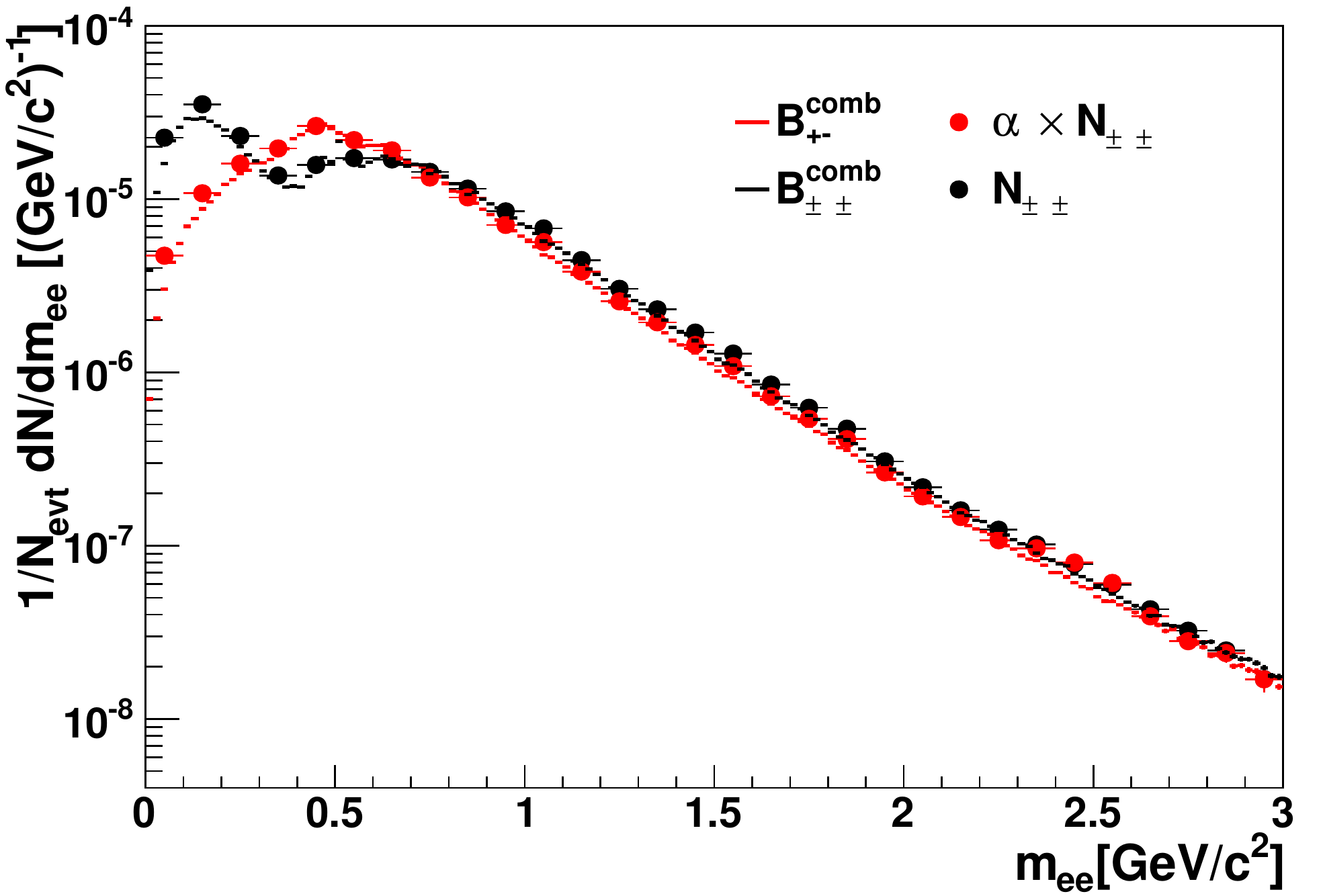}
\caption{\label{Fig:rel_accp}
Mass distribution for the combinatorial background determined by
event mixing \Bunlike\ and \Blike as the red and black line,
respectively. The shape difference due to the difference in
acceptance between like-sign and unlike-sign pairs in PHENIX is
clearly visible. Also shown are foreground like-sign pairs \Nlike
(black points) and \Nlike corrected for the acceptance difference
(red points). The differences between points and lines are the the
correlated background.
}
\end{figure}

There are two different approaches to estimating the background, (i) 
the like-sign subtraction technique based on the measured like-sign 
foreground \Nlike\ or (ii) the event mixing technique. In the PHENIX 
experiment the acceptance for like and unlike-sign pairs is different 
due to the two arm geometry and thus the shape of the invariant mass 
distributions are different as illustrated in 
Figure~\ref{Fig:rel_accp}. We therefore traditionally have used the 
event mixing technique. In this method, combinatorial background is 
estimated by taking an electron from event {\it i} and pairing it 
with a positron from event {\it j}($\neq i$). This is a powerful 
approach as it allows for an extremely high statistics estimation of 
the background \cite{ppg088}.  However, such an estimation must be 
normalized with a precision much better than the $S/B$. In addition, 
the mixed event spectra do not contain any of the correlated 
background and therefore these additional pairs must be estimated 
using Monte Carlo methods.

In this paper we use the like-sign subtraction technique, which 
avoids the complications inherent in the mixed event background 
estimation. This method uses the acceptance difference for like and 
unlike-sign pairs, described in Section \ref{sec:likesign}.

\subsubsection{Correlated Background}
\label{sec:corr_bkg}

There are two sources of correlated background: {\it cross pairs} and 
{\it jet pairs} \cite{ppg085}. Cross pairs are correlated through a 
hadron decay that results in two $e^+e^-$ pairs.  These pairs 
originate from $\pi^0$ and $\eta^0$ double-Dalitz decays 
($\pi^{0}(\eta)\rightarrow\gamma^{*}\gamma^{*}{\rightarrow}e^{+}e^{-}e^{+}e^{-}$), 
a single-Dalitz decay accompanied by a photon conversion 
($\pi^{0}(\eta)\rightarrow\gamma\gamma^{*}{\rightarrow}e^{+}e^{-}e^{+}e^{-}$), and
diphoton decays with both photons converting ($\pi^{0}(\eta) 
\rightarrow \gamma\gamma \rightarrow e^{+}e^{-}e^{+}e^{-}$).  The 
cross pair correlation arises because of the small opening angle 
between the virtual and/or real decay photons.  The resulting 
dielectrons tend to manifest at low mass and high $p_T$.

Jet pairs are the other major source of correlated \ee background. In this 
case, the electron and positron are decay products of different hadrons 
inside jets. Di-jet production and fragmentation causes a correlation in 
the parent hadrons, which is inherited by the daughter electrons. When the 
electron and positron are from opposing (back-to-back) jets, the pair 
typically has low $p_T$ and high mass. When they arise from two hadrons in 
the same jet, the pair typically has a high $p_T$ and low mass.

Since cross pairs and jet pairs result from two \ee pairs, correlated 
pairs with like and unlike-sign are produced at the same rate.  This 
fact can be exploited to correct for correlated background in the 
unlike-sign distribution.

\subsubsection{Like-sign Subtraction}

\label{sec:likesign}

The like-sign subtraction technique uses the foreground like-sign 
pairs \Nlike\ to determine the background. This has two distinct 
advantages over the event mixing technique. First, the measured yield 
\Nlike\ requires no additional absolute normalization.  The second 
advantage, which was mentioned in the previous section, is that 
\Nlike\ contains the identical amount of correlated background as the 
measured $e^+e^-$ pairs \Nunlike. Hence, no independent simulation of 
the correlated background is needed.

This method, however, can be used in PHENIX only after correcting for 
the different acceptance for like-sign and unlike-sign pairs of the 
two-arm configuration (see Fig.~\ref{Fig:rel_accp}). This 
correction is provided by the ratio of the acceptance functions for 
unlike- and like-sign pairs, the relative acceptance correction, 
$\alpha$, which is due solely to the detector geometry and is 
determined using mixed events as follows:

\begin{eqnarray}
\alpha(m,p_T)&=&\frac{\Bunlike(m,p_T)}{\Blike(m,p_T)}\label{eq_alpha}.
\end{eqnarray}
The ratio of mixed-event unlike-sign to like-sign pairs is calculated 
differentially in mass and $p_{T}$ and is applied to each run group 
separately.

Figure~\ref{Fig:rel_accp} shows the mass distribution for the unlike 
and like-sign pairs in mixed events, \Bunlike and \Blike, 
respectively. Also shown is the mass spectrum for like-sign pairs 
\Nlike. The relative acceptance correction translates \Nlike to the 
unlike-sign pair space via $\Nunlike = \alpha \times \Nlike$. 
Deviations between the $\alpha$ corrected like-sign spectrum and the 
unlike-sign mixed events correspond to the cross pairs and jet pairs.

\begin{figure}[htb]
\includegraphics[width=0.95\linewidth]{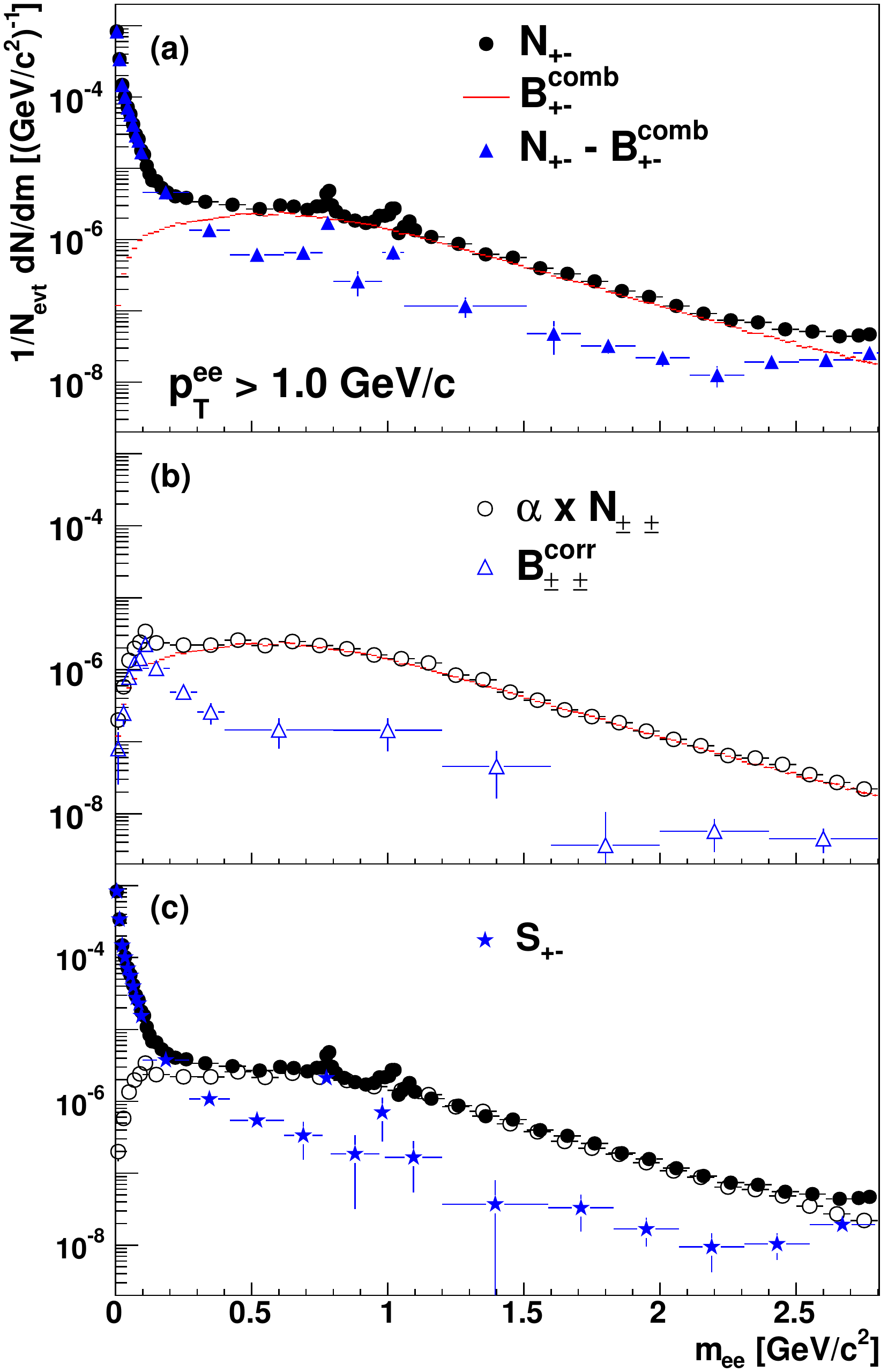}
\caption{\label{Fig:signal_bkg}
The top panel shows the $e^+e^-$ pair foreground $N_{+-}$, the 
combinatorial background \Bunlike\ determined through event mixing, 
and the difference of the two which is the sum of the signal we are 
interested in \signal\ and the correlated background \Bcorr\ that 
still needs to be subtracted. Shown in the middle panel is the 
estimate of the correlated background \Bcorr, which is the difference 
between the foreground like-sign pairs \Nlike\ corrected for the 
relative acceptance difference $\alpha$ between \Nlike and \Nunlike 
(see Fig.~\ref{Fig:rel_accp} and Eq.~\ref{eq_alpha}) and the 
combinatorial background \Bunlike. The bottom panel shows the signal 
\signal\ which is calculated as $\Nunlike - \alpha \times \Nlike$. In 
this plot, the combinatorial background is normalized in a region 
with minimal correlated background\cite{ppg085}.
}
\end{figure}

The subtraction procedure is illustrated in 
Fig.~\ref{Fig:signal_bkg}.  It illustrates the steps to transform 
the measured $e^+e^-$ pairs \Nunlike in 
Fig.~\ref{Fig:signal_bkg}(a) to the signal of interest \signal\ in 
Fig.~\ref{Fig:signal_bkg}(c).  Figure~\ref{Fig:signal_bkg}(a) shows 
\Nunlike, \Bunlike and their difference, which corresponds to the 
signal \signal plus the correlated background \Bcorr. The middle 
panel of Fig.~\ref{Fig:signal_bkg}(b) shows \Bcorr\ calculated as the 
difference, $\alpha \times \Nlike ~- ~\Bunlike$. The signal \signal\ 
is given in Fig.~\ref{Fig:signal_bkg}(c). The actual background 
subtraction is done double-differentially, and separately for each 
run group, as well as separately for minimum bias and electron 
triggered events.
\begin{eqnarray} 
\signal(m,p_T) = \Nunlike(m,p_T) - 
\Bunlike - \Bcorr \nonumber \\ = \Nunlike(m,p_T) - \alpha(m,p_T) 
\times \Nlike(m,p_T). 
\end{eqnarray}

For the electron triggered events, the trigger used in the data 
collection biases the single electron distribution towards high \pt 
and as such the triggered events can not be mixed with each other. 
Thus to generate the correct combinatorial background shape of \ee 
pairs, the mixed events are generated from the minimum bias data 
sample, but as in the real events, they are required to satisfy the 
trigger requirement. Every mixed pair therefore contains at least one 
electron that fulfills the trigger condition\cite{phd_kamin}.

\subsubsection{$B$ meson decay chains}
\label{sec:oscillations}

Approximately 1/3 of the $ee$ pairs from \bb production are like-sign 
pairs. Using the like-sign subtraction technique, these are removed 
from the signal \signal. The main decay chains for $B$ and $D$ mesons 
to $ee$ pairs are shown in Table.~\ref{Tab:decays}. While for \cc, 
only the direct semi-leptonic decays, (1) in Table.~\ref{Tab:decays}, 
contribute, many more possibilities exist for \bb. Decay combinations 
(1),(1) and (2),(2) lead to $e^+e^-$ pairs, while combinations (2)(1) 
and (1)(2) lead to $e^-e^-$ and $e^+e^+$ pairs due to the flavor 
change in the decay. The last decay chain (3) involves decay of a 
single $b$ or $\bar{b}$ and produces only \ee pairs. Since the 
semi-leptonic decay channels for $B$ and $D$ mesons have 
approximately equal branching ratios, and more than 90\% of $B$ 
mesons decay to $D$, all three groups of decays are approximately 
equally likely. This results in about a third of all $ee$ pairs from 
\bb decays being like-sign pairs.

\begin{table}[htb]
\caption{Summary of the most relevant \cc and \bb decay chains that 
contribute to \ee pairs. The effective branching ratio averages over 
all possible meson combinations.} {\label{Tab:decays}}
\begin{ruledtabular} \begin{tabular}{ccccc}
&    Mode   & Decay chain & Effective B.R. & \\
    \hline
&    (1)    & $D \to e^+ X$                    & 9.4\% & \\
&    (1)    & $B \to e^+ X$                    & 11\%   & \\
&    (2)    & $B \to \bar{D} X \to e^- X$      & 8.5\%  & \\
&    (3)    & $B \to \bar{D} e^+ X \to e^+e^- X$ & 0.8\% & \\
\end{tabular} \end{ruledtabular}
\end{table}

Another important difference between $ee$ pair production from \bb 
compared to \cc is that particle-antiparticle oscillations between 
$B^{0}$ and $\bar{B}^{0}$ can change one of the charges in an $ee$ 
pair \cite{pbm_osc}. A $B^{0}_{d}$ oscillates with a probability of 
$\sim$17\% while a $B^{0}_{s}$ does so $\approx$49\% of the time 
\cite{B_mixing}. Therefore, in the all decay chain combinations 
involving (1) or (2) from Table~\ref{Tab:decays}, there is ~20\% 
probability for a sign change.

It is thus vital to treat the simulations with the same procedure as 
the data in order to properly account for all of the heavy flavor 
pairs.  Both \pythia~\cite{pythia} and Monte Carlo at 
next-to-leading-order (\mcnlo)~\cite{mcnlo_main} calculations 
generate the proper like-sign yield from heavy flavor sources. As in 
the data analysis, we subtract this like-sign contribution from the 
unlike-sign yield in the simulations. Only then are comparisons made 
to the data.

\subsection{Efficiency Corrections}\label{sec:efficiency}

The \ee signal \signal\ is corrected for single particle detection 
and identification efficiency to obtain the dielectron yield in the 
PHENIX acceptance:

\begin{eqnarray}\nonumber
\frac{d^2N}{dm_{ee}dp_{T}^{ee}} 
&=& \frac{1}{N_{\rm evt}^{\sc sampled}} 
\cdot \frac{1}{\Delta m_{ee}} 
\cdot \frac{1}{\Delta p_{T}^{ee}}
\cdot \frac{1}{\varepsilon_{\rm rec}(m,p_T)}\\
&&\cdot \frac{1} {\varepsilon_{\rm ERT} (m, p_T)}
\cdot \signal(m,p_T) \cdot C_{\rm bias}.
\end{eqnarray}
The reconstruction efficiency $\varepsilon_{\rm rec}(m,p_T)$ is 
evaluated using a \geant Monte Carlo simulation of the PHENIX 
detector. It accounts for losses in yield due to  dead areas in the 
detector, track reconstruction efficiency, single track quality 
cuts, electron identification cuts, and \ee pair cuts. Since the 
detector performance varies from run group to run group, efficiency  
is evaluated separately for each run group. The inverse 
$(\varepsilon_{\rm rec}(m,p_T))^{-1}$ is used to correct the 
\signal\ to represent the yield in the ideal PHENIX 
acceptance\footnote{The PHENIX acceptance is parameterized as 
function of the azimuthal angle $\phi$ of a track, its \pt, and 
charge sign q by conditions for the DC and the RICH for each 
spectrometer arm separately: $\phi_{\rm min} < \phi + q k_{\rm DC}/\pt < 
\phi_{\rm max}$ and $\phi_{\rm min} < \phi + q k_{\rm RICH}/\pt < \phi_{\rm max}$.
The parameters are $k_{\rm DC}=0.206$ rad GeV/$c$, $k_{\rm RICH}=0.309$ rad 
GeV/$c$, $\phi_{\rm min}= -3/16 \pi$ to $\phi_{\rm max}=5/16 \pi$, 
and$\phi_{\rm min}= 11/16 \pi$ to $\phi_{\rm max}=19/16 \pi$.}.
No correction is made for pair acceptance, as 
the magnitude of such corrections depends upon the pair production 
process and thus the opening angle between the electron and 
positron. 

The trigger efficiency $\varepsilon_{\rm ERT}(m,p_T)$ for \ee pairs 
is measured by requiring that one of the electrons in the pair 
satisfies the single electron trigger conditions. The resulting mass 
spectrum is compared to that from minimum bias events to evaluate the 
trigger efficiency. The trigger approaches full efficiency for pair 
masses above approximately 2 GeV/$c^2$.

The factor $C_{\rm bias}=0.889\pm 0.003$ accounts for the correlation 
between heavy flavor events and an increase in the charge deposited 
in the BBC \cite{ppg160} as well as any inefficiency in the BBC 
trigger.  It is calculated in a Glauber Monte Carlo-based framework 
that includes the BBC response. The corrected yield represents the 
heavy flavor yield corresponding to the inelastic \dA cross section 
of $\sigma_{\rm inel}^{d\rm{Au}} = 2.3 \pm 0.1 ~b$ \cite{ppg160}.

\subsection{Systematic Uncertainties}

The systematic uncertainties on the \ee yield arise from 
uncertainties on the dielectron reconstruction efficiency, the single 
electron trigger efficiency, and the precision of the background 
determination.

The uncertainty on electron reconstruction is based on the 
reproducibility of the final result using multiple cut variations 
both on single electrons and on electron pairs. The cuts varied 
include electron identification, conversion rejection, and pair cuts.  
The conversion rejection and pair cuts are less influential and only 
affect the low mass region ($<600$~MeV/$c^{2}$).  The uncertainties 
are evaluated by reconstructing simulated dielectrons using a full 
\geant\ Monte Carlo simulation of the PHENIX detector.  Detector 
dead areas can vary slightly within a given performance-based run 
group.  Typical run-by-run variations were analyzed in addition to 
group-by-group variations, in order to evaluate the systematic 
uncertainties from detector performance.  In the intermediate (1-3 
GeV/$c^2$) and high mass regions ($>$3 GeV/$c^2$), these 
uncertainties vary between 10-20\%.

The precision of the trigger efficiency correction depends on the 
available statistics in the minimum bias data sample as well as on 
the super module segmentation of the EMCal. An EMCal super module is 
a group of 12$\times$ 12 (or 6 $\times$ 4) lead-scintillator  
towers\cite{emcNIM}.  The trigger efficiency is calculated using the 
statistically independent minimum bias data for each EMCal super 
module separately within each run group.  The single electron trigger 
efficiencies are then used in the simulation to obtain pair trigger 
efficiency.  The triggered data is used above pair 
$m_{T}>1.5$~GeV/$c$ and contributes only a 5\% uncertainty to the 
final result.

The dominant source of systematic uncertainty is the accuracy of the 
relative acceptance correction.  Since it is a mass and $p_{T}$ 
dependent scale factor applied directly to the background, it affects 
the overall uncertainty in proportion to the background-to-signal 
ratio.  This correction is very sensitive to the fluctuations in 
detector dead area that exist within a run group.  Dedicated Monte 
Carlo simulations were performed to determine the effect of removing 
or including various regions of the PHENIX central arms.  These 
regions were chosen to reflect realistic geometry including EMCal 
modules/super modules, DC wires grouped by power input and 
signal output, and shifted positions of intrusive support structures.  
This uncertainty ranges from $<$5\% at high mass ($>$5 GeV/$c^2$) to 
$\sim$25\% below 2.5~GeV/$c^{2}$.

Table~\ref{Tab:syst_errs} summarizes the magnitude of the systematic 
uncertainty arising from various sources and the affected mass 
ranges.

\begin{table}[htb]
\caption[Systematic Uncertainties]{\label{Tab:syst_errs} Systematic 
uncertainties of the dilepton yield due to different sources with an 
indication of the applicable mass range. The transverse mass is 
defined as $m_T = \sqrt{m^2+p_T^2}$.
}

\begin{ruledtabular} \begin{tabular}{ccc}
 Component & Syst.uncertainty & Mass (GeV/$c^{2}$)\\
    \hline
    Pair reconstruction      &  14\%   & 0--14\\
    Conversion rejection     &  6\%    & 0--0.6\\
    	                     &  0\%   & $>$0.6\\
    Pair cuts                &  5\% & 0.4--0.6\\
    Trigger efficiency           &  5\% & $m_{T} \geq 1.5$\\
    Dead area, run groups&  15\%   & 0--2.5\\
    ~                        &  10\%   & 2.5--14\\
    Relative acceptance  &  5\%$\times B/S$ & 0--2.5\\
    ~                        &  2\%$\times B/S$ & 2.5--5\\
    ~                        &  1\%$\times B/S$ & $>$5\\
     \end{tabular} \end{ruledtabular}
\end{table}

\section{Results}

\subsection{Yield of $e^+e^-$ pairs}  

\begin{figure*}
\includegraphics[width=1.0\linewidth]{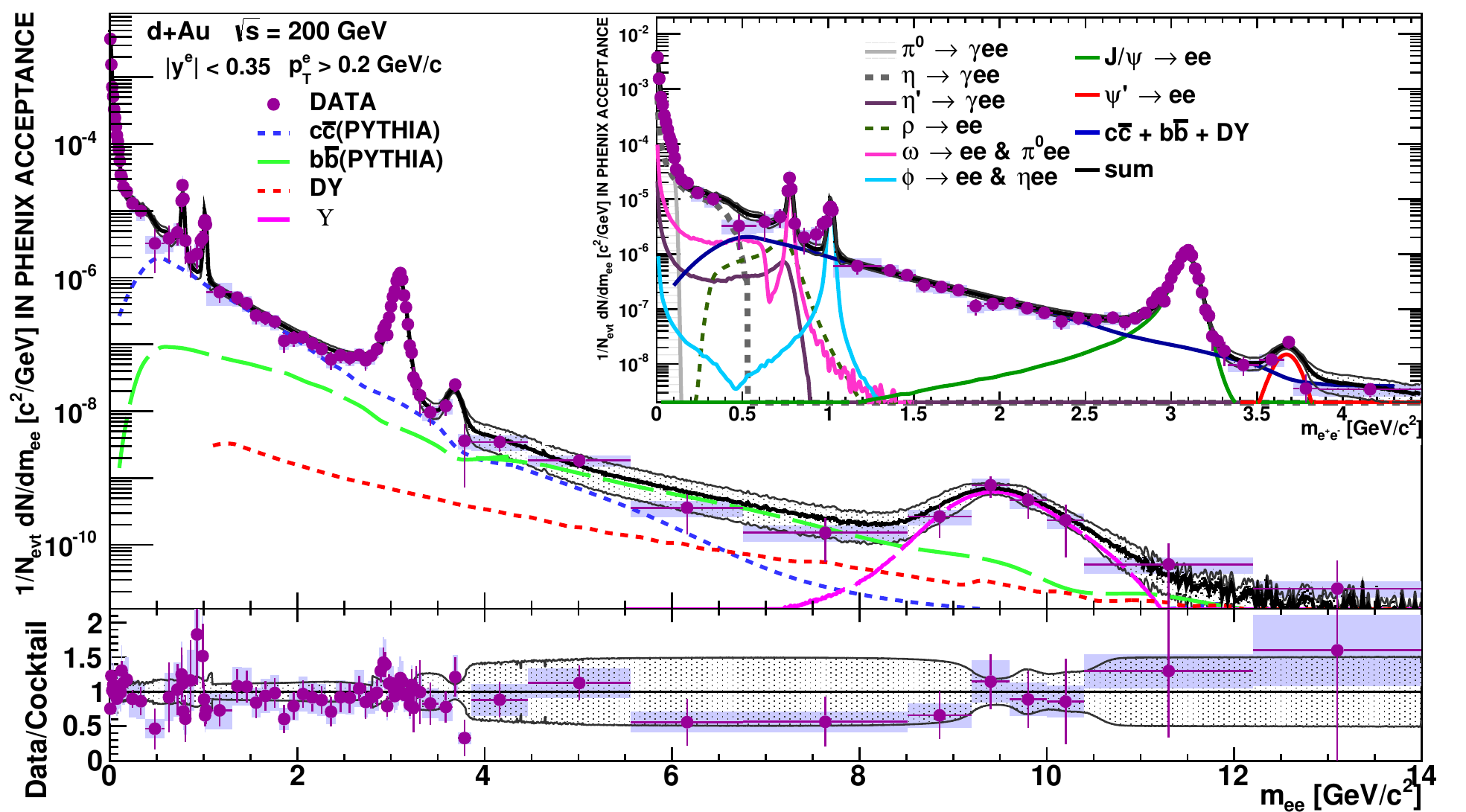}
\caption{\label{Fig:dAu_mass_spectrum}
Inclusive \ee pair yield from minimum bias \dA collisions as a 
function of mass. The data are compared to our model of expected 
sources. The inset shows in detail the mass range up to 4.5 
GeV/$c$$^2$. In the lower panel, the ratio of data to expected 
sources is shown with systematic uncertainties.
} 
\end{figure*}

Figure~\ref{Fig:dAu_mass_spectrum} shows the mass projection of the 
measured double differential \ee pair yield in the ideal PHENIX acceptance 
(as described in footnote$^1$).  The inset shows the mass spectrum up to 
4.5 GeV/$c^2$, and a detailed cocktail of hadronic decay sources that 
contribute to the mass spectrum below 4.5 GeV/$c^2$. The main figure shows 
the mass distributions of charm, bottom and Drell-Yan \ee pairs obtained 
using \pythia.  One can clearly see that the resonances lie atop a 
continuum, which is dominated by three body decays of pseudoscalar and 
vector mesons for masses below 1.0 GeV/$c$$^2$. Above 1.0 GeV/$c$$^2$ the 
continuum is dominated by pairs from semi-leptonic decays of heavy flavor, 
with the bottom contribution becoming more important at higher mass.

The lower panel of the Figure~\ref{Fig:dAu_mass_spectrum} shows the 
ratio of data to the expected sources. The shape of the measured mass 
spectrum is well described by the expected sources over the entire 
mass range. For the mass range below 1.0~GeV/$c^2$, the cocktail is 
absolutely normalized and shows a good agreement to the data. For the 
high mass region, the \ee pair continuum from heavy flavor decays is 
normalized to the data to extract the bottom and charm cross section 
as discussed below.

\subsection{Expected sources of $e^+e^-$ pairs}  

Many sources contribute to the inclusive $\ee$ pair yield, so an 
in-depth understanding of the expected sources and their double 
differential distribution in $\ee$ pair mass and \pt is necessary to 
interpret the data. We use the detailed component-by-component 
simulation developed in \cite{ppg088}, as a benchmark. The cocktail 
includes pseudoscalar and vector meson decays, semi-leptonic decays 
of heavy flavor, and \ee pairs created through the Drell-Yan 
mechanism.

The pseudoscalar mesons, $\pi^0$ and $\eta $, and vector mesons, 
$\omega$, $\phi$, $J/\psi$ and the $\Upsilon$, are generated based on 
measured differential \dA cross sections \cite{phenix_pi_neutral, 
phenix_pi_charged, phenix_omega, phd_deepali, phenix_jpsi, ppg142}. 
The contributions from mesons not directly measured in $\dA$ 
($\eta'$, $\rho$, and $\psi'$) are determined relative to the 
measured mesons ($\eta$, $\omega$, $J/\psi$, respectively) using 
particle ratios from \pp or jet fragmentation \cite{ppg085}. Decay 
kinematics, branching ratios, electromagnetic transition form 
factors, etc. are based on the most up-to-date information from the 
Particle Data Group \cite{PDG}.
The yield of \ee pairs created through the Drell-Yan mechanism was 
simulated using {\sc pythia}\footnote{Drell-Yan 
{\sc pythia}-6~\protect\cite{pythia}, using parameters: MSEL=0, 
MSTP(43)=3, MSTP(33)=1, MSTP(32)=1, MSUB(1)=1, MSTP(52)=2, MSTP(54)=2, 
MSTP(56)=2, MSTP(51)=10041 (CTEQ6LL), MSTP(91)=1, PARP(91)=1.5, 
MSTP(33)=1, MSTP(31)=1.38, MSTP(32)=4, CKIN(3)=0.5, CKIN(1)=0.5, 
CKIN(2)=-1.0, CKIN(4)=-1.0, MSTP(71)=0 }
For the normalization we use a cross section of $34\pm28$~nb, which was 
determined by a simultaneous fit of the data at high mass to Drell-Yan, 
charm, and bottom contributions using the \pythia simulation. The 
systematic uncertainty in the Drell-Yan cross section is propagated 
through the subsequent heavy flavor cross section analysis. This 
uncertainty has a negligible effect ($<5\%$) on the final result of the 
bottom cross section. As can be seen from 
Fig.~\ref{Fig:dAu_mass_spectrum}, the contribution from Drell-Yan is 
extremely small below $\approx$ 5 GeV/$c^2$. It remains a minor 
contribution to the dielectron pair spectrum below 10 GeV/$c^2$.
 
The double differential contribution from semi-leptonic decays of 
heavy flavor are simulated using two different \pp event generators, 
\pythia and \mcnlo. The cross sections for \cc and \bb in the 
cocktail shown in Fig.~\ref{Fig:dAu_mass_spectrum} are the ones 
extracted from this work, as discussed below.

The \pythia program generates heavy quark pairs by calculating the leading 
order pQCD gluon fusion contributions.  We used \pythia in forced \cc or 
\bb production 
mode\footnote{Heavy flavor {\sc pythia}-6~\protect\cite{pythia}, using 
parameters MSEL=4 (\cc) or 5 (\bb), MSTP(91)=1, PARP(91)=1.5, MSTP(33)=1, 
PARP(31)=1.0, MSTP(32)=4, PMAS(4)=1.25, PMAS(5)=4.1"} to match 
Ref.~\cite{ppg085}, and CTEQ5L as the input parton distribution function.

The \mcnlo package (v. 4.03) \cite{mcnlo_main, mcnlo_qq} is an NLO 
simulation that generates hard scattering events to be passed to 
\herwig (vers. 6.520) \cite{herwig} for fragmentation into the 
vacuum.  Since the package is a two-step procedure consisting of 
event generation and then fragmentation, care is taken to pass the 
color flow of each parton configuration from the generator to 
\herwig.  In addition, since flavor creation (i.e., $qq \rightarrow 
QQ$ and $gg \rightarrow QQ$) processes at order $\alpha_S^2$ can 
generate some of the higher order processes through parton showering, 
\mcnlo keeps track of this to ensure an accurate result.  While the 
default \mcnlo package generates \bb events, it does not incorporate 
\cc events.  Thus, we altered the default package to enable charm 
production\footnote{This trivial adaptation was reviewed by the 
original \mcnlo authors via private communication.}.
Because both \mcnlo 
and \herwig use the standard PDG process ID codes \cite{PDG}, we 
changed the process code from -1705 ($H_{1}H_{2} \rightarrow 
b\bar{b}+X$) to -1704 ($H_{1}H_{2} \rightarrow c\bar{c}+X$) and 
adjusted the heavy quark mass to the charm quark, $1.29$~GeV/$c^{2}$.  
No other parameters were modified. In contrast to \pythia, the 
running parameters of \mcnlo does not need to be fine-tuned for 
different analyses.  CTEQ6M \cite{lhapdf} was used to provide the 
input parton distribution function.

The electrons and positrons from all simulations are filtered through 
the PHENIX acceptance \cite{ppg088}. The \ee pair acceptance depends 
on the production process, which determines the correlation between 
the electron and positron.  For pseudoscalar and vector meson decays, 
the \ee pairs originate from an intermediate virtual photon that 
correlates the momenta of $e^+$ and $e^-$. For \ee pairs from heavy 
flavor decays the correlation is governed by the interplay of two 
contributions: (i) the QCD production of the \qq pair, which 
determines the rapidity distribution of the pair, the rapidity gap 
between $q$ and $\bar{q}$ and the extent to which they are 
back-to-back in azimuthal angle; and (ii) the decay kinematics of the 
two independent semi-leptonic decays. The latter tends to randomize 
the correlation if the mass of the quark is large compared to its 
momentum. In the limit of very large quark masses the decays will 
occur at rest and the $e^+$ and $e^-$ momenta will be determined 
exclusively by the independent decays. In contrast, for small quark 
masses the decay products will be boosted along the momenta of the 
parent quarks and thus their correlation will closely reflect the 
correlations between the parent quarks.

The differences between the acceptance for \ee pairs from charm and 
bottom production are documented in Tables~\ref{Tab:cc_pairs} to 
\ref{Tab:bb_ee_pairs}. While only 1 out of 500 \ee pairs from charm 
production is accepted in PHENIX, 1 out of 120 pairs from bottom 
production is accepted. This can be compared to the limiting case of 
very large quark masses, for which the direction of the decay $e^+$ 
and $e^-$ are independent and approximately 1 of 80 \ee pairs will 
fall into the PHENIX acceptance. The acceptance for \ee pairs from 
\bb is only 30\% different from this limiting case, while for \cc the 
deviation is more than a factor of five. This suggests that the 
acceptance for pairs from \bb is driven mostly by decay kinematics, 
and thus depends only a little on the correlation between the $b$ and 
$\bar{b}$. Consequently the model dependence must be much smaller for 
\bb than for \cc.
 
Comparing \pythia and \mcnlo in Table~\ref{Tab:cc_ee_pairs} and 
Table~\ref{Tab:bb_ee_pairs} shows that indeed the difference between 
the acceptance calculated with \pythia and \mcnlo is much 
smaller for \bb than for \cc pairs.  For bottom production the 
difference is about 5\%, while in the charm case the acceptance is 
different by a factor of 1.2, which increases to 2.2, if one restricts 
the mass range to above 1.16 GeV/$c$$^2$.  Most of this 
model-dependence is already apparent when going from 4$\pi$ to a 
restricted rapidity coverage of $\Delta y =1$ for $e^+$ and $e^-$, 
and does not significantly increase when restricting to the smaller 
PHENIX aperture.

\begin{table}
\caption{Number of \cc pairs at midrapidity in $y_{\cc}=1$ and 
$y_{\cc}=0.7$ relative to $4\pi$. $y_{\cc}$ corresponds to the 
rapidity of center-of-mass of \cc pair.
}
    \begin{ruledtabular} \begin{tabular}{ccccc}
&      Acceptance    & \pythia \cc pairs & \mcnlo \cc pairs & \\
      \hline
&      4$\pi$                               &  1      &    1 & \\
&      $|y_{c\bar{c}} |< 0.5$       &  0.275  &  0.297 & \\
&      $|y_{c\bar{c}} |< 0.35$      & 0.2   &  0.215 & \\
    \end{tabular} \end{ruledtabular}
    \label{Tab:cc_pairs}
\end{table}

\begin{table}
\caption{Yields of \ee pairs from \cc, measured in units of one \cc 
pair per event divided by the effective semi-leptonic branching ratio 
squared ${F^ {c\bar{c}} _{BR}} = {(B.R.(c\rightarrow e ))}^2$, where 
B.R. is the effective branching ratio of 9.4\%.}
    \begin{ruledtabular} \begin{tabular}{ccc}
      Acceptance & \pythia \ee pairs   & \mcnlo \ee pairs     \\
      & from \cc [${F^ {c\bar{c}} _{BR}}^{-1}$] 
      & from \cc [${F^ {c\bar{c}} _{BR}}^{-1}$]\\
      \hline
      4$\pi$     & 1                    &  1                     \\
      $|y_{e^+} \& y_{e^-} |< 0.5$  & 0.042   & 0.035            \\
      $|y_{e^+} \& y_{e^-} |< 0.5 ~\&\&$  & 0.0047   & 0.00022         \\      
      $m_{e^+e^-}>1.16 $GeV/$c^2$            & &            \\        
      $ |y_{e^+} \& y_{e^-} |< 0.35$     & 0.021   &  0.017    \\
      $ |y_{e^+} \& y_{e^-} |_{\rm PHENIX}$  & 0.0023  & 0.0016      \\
      $ |y_{e^+} \& y_{e^-} |_{\rm PHENIX} ~\&\&$  & 0.00044  & 0.0002   \\
      $ m_{e^+e^-}>1.16 $GeV/$c^2$            & &            \\  
    \end{tabular} \end{ruledtabular}
\label{Tab:cc_ee_pairs}
\end{table}

\begin{table}
\caption{Number of \bb pairs at midrapidity in $y_{\bb}=1$ and 
$y_{\bb}=0.7$ relative to $4\pi$. $y_{\bb}$ corresponds to the 
rapidity of center-of-mass of \bb pair.}
    \begin{ruledtabular} \begin{tabular}{ccccc}
&      Acceptance & \pythia \bb pairs & MC@NLO \bb pairs & \\
      \hline
&      4$\pi$                               &  1   &    1 & \\
&      $|y_{b\bar{b}} |< 0.5$       &  0.39  &  0.40 & \\ 
&      $|y_{b\bar{b}} |< 0.35$      & 0.28   &  0.29 & \\ 
&      
    \end{tabular} \end{ruledtabular}
    \label{Tab:bb_pairs}
\end{table}

\begin{table}
\caption{Yields of \ee pairs from \bb, measured in units of one \bb 
pair per event divided by the effective semi-leptonic branching ratio 
squared ${F^ {b\bar{b}} _{BR}} = {(B.R.(b\rightarrow e ))}^2$, where 
B.R. is the effective branching ratio of 15.8\% using a like-sign 
pair subtraction, or 22\% not considering the like-sign pairs.
    }
    \begin{ruledtabular} \begin{tabular}{ccc}
      Acceptance & \pythia \ee pairs    & \mcnlo \ee pairs     \\
      & from \bb [${F^ {b\bar{b}} _{BR}}^{-1}$] 
      & from \bb [${F^ {b\bar{b}} _{BR}}^{-1}$]\\
      \hline
      4$\pi$     & 1                    &  1                     \\
      $|y_{e^+} \& y_{e^-} |< 0.5$       & 0.095   & 0.091    \\
      $|y_{e^+} \& y_{e^-} |< 0.5$       & 0.0425   & 0.0395   \\
      $ m_{e^+e^-}>1.16 $GeV/$c^2$         &         &           \\
\\
      $ |y_{e^+} \& y_{e^-} |< 0.35$     & 0.048   &  0.046   \\
      $ |y_{e^+} \& y_{e^-} |_{\rm PHENIX}$  & 0.0084  & 0.0080   \\
      $ |y_{e^+} \& y_{e^-} |_{\rm PHENIX}$  & 0.00368  & 0.0037   \\
      $ m_{e^+e^-}>1.16 $GeV/$c^2$            & &                \\  
    \end{tabular} \end{ruledtabular}
    \label{Tab:bb_ee_pairs}
  \end{table}

The correlations of the $q$ and $\bar{q}$ are very different in 
\pythia and \mcnlo. While in \mcnlo the correlation is due to 
including NLO terms explicitly in the pQCD calculation, in the first 
order \pythia calculation the correlation is largely determined by 
the specific implementation of intrinsic transverse momentum ($k_T$). 
While both models predict similar momentum distributions for the 
individual $q$ and $\bar{q}$, the opening angle distributions for the 
\qq pairs are different and thus the mass distributions in 4$\pi$ 
differ substantially. These differences decrease upon selecting decay 
\ee pairs that fall in the PHENIX acceptance, so the shape of the 
mass and \pt distributions from the two models are quite similar. 
Thus in the PHENIX acceptance, the model differences in the \qq 
correlations surface mostly through different fractions of \ee pairs 
that fall in the acceptance.

For \bb pairs the decay kinematics have a different effect than for 
\cc. About 50\% of the \ee pairs from \bb production involve only the 
decay of the $b$ or $\bar{b}$ quark through the decay chain (3) from 
Table~\ref{Tab:decays} and thus are {\it a priori} insensitive to the 
opening angle of the \bb pair.

Since more than 90\% of the $B$-mesons have momenta much smaller than 
their mass, the decay electron is less likely to move in the same 
direction as the parent meson. Consequently the correlation between 
$e^+$ and $e^-$ from decays of $b$ {\it and} $\bar{b}$ through decay 
chains (1) and (2) in Table~\ref{Tab:decays} is smeared. The fraction 
of \ee pairs in our acceptance from \bb is much less sensitive to the 
correlations between the $b$ and $\bar{b}$. We have tested this 
conclusion by randomizing the correlation between $b$ and $\bar{b}$ 
and found that the acceptance remains unchanged while there is a 
significant difference for \cc.

Since the acceptance of \ee pairs from \bb is mostly driven by decay 
kinematics and not by the model dependent production mechanism, the 
fraction of \ee pairs must also be less sensitive to any 
cold-nuclear-matter effects that alter the $b$ or $\bar{b}$ after 
they are produced. For the lighter \cc quarks the sensitivity to the 
opening angle between the $c$ and $\bar{c}$ is much larger, implying 
larger model dependence and consequently cold-nuclear-matter effects 
may have a larger influence on the distribution of dielectrons from 
\cc. The results obtained in this analysis seem also insensitive to 
nuclear modifications of the parton distribution function; when using 
EPS09 \cite{EPS09} for the \mcnlo or \pythia calculation the 
acceptance factor for \ee pairs from \bb and \cc production change by 
less than 5\%.

The simulated \ee pairs are folded with the experimental momentum 
resolution as well as with the energy loss due to bremsstrahlung. As 
a result we obtain the double differential \ee pair yield for the 
expected sources that can be directly compared to the measured yield. 
All components are absolutely normalized, except for the heavy 
flavor contributions, which are used to determine the bottom and 
charm cross section from the \ee pair data, and the Drell-Yan 
contribution, which is negligibly small and was fixed to be consistent 
with the data.

\subsection{\ee pairs from heavy flavor decays}\label{results_hf}

\begin{figure}[htb]
  \includegraphics[width=1.0\linewidth]{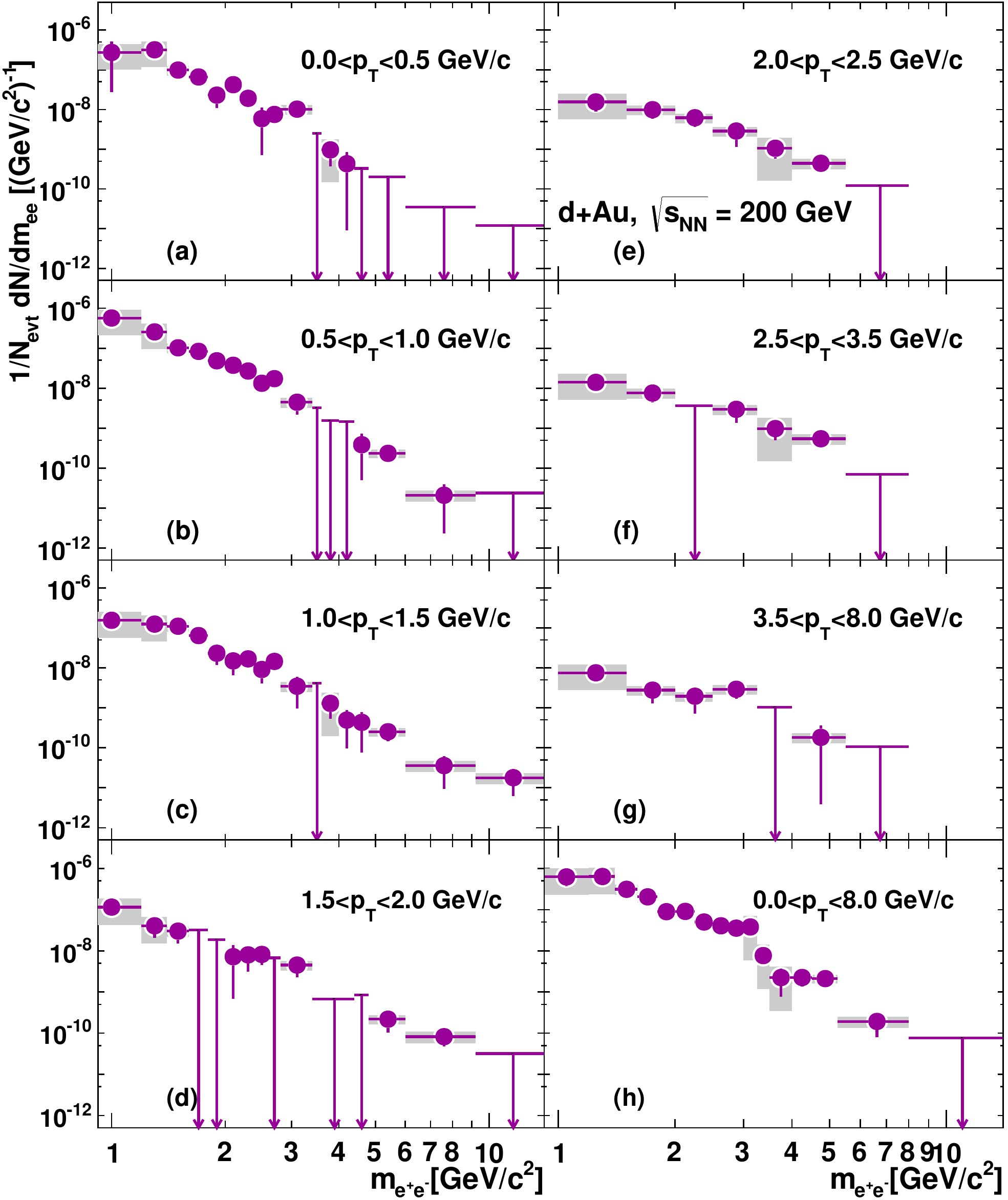}
\caption{Double differential \ee pair yield from semi-leptonic decays 
of heavy flavor in inelastic \dA collisions. Shown are mass 
projections in slices of \pt. The \pt intervals are indicated in each 
panel. Systematic uncertainties are shown as bars, downward pointing 
arrows indicate upper limits at 90\% CL.
}
\label{Fig:doublediffdata}
\end{figure}

In order to access the heavy flavor yield, we subtract the yield of 
the pseudoscalar and vector mesons as well as the Drell-Yan 
contribution from the measured dielectron spectra. The subtraction is 
done double differentially in mass and \pt. The results are shown in 
Fig.~\ref{Fig:doublediffdata} as mass spectra in slices of 
transverse momentum. The data are plotted above 1.0 GeV/$c$$^2$, as 
lower mass \ee are dominated by hadronic decay contributions. In the 
mass regions where the inclusive \ee yield is dominated by vector 
meson decays only upper limits can be quoted for the subtracted 
spectra.  We use \pt bins of 500 MeV/$c$ up to \pt =3 GeV/$c$. 
Above $\pt=3.0$ GeV/$c$, statistical limitations dictate the use of 
broader \pt bins.

\section{Heavy Flavor Cross Section Determination}\label{xsec_det}

\begin{figure}[htb]
\includegraphics[width=1.0\linewidth]{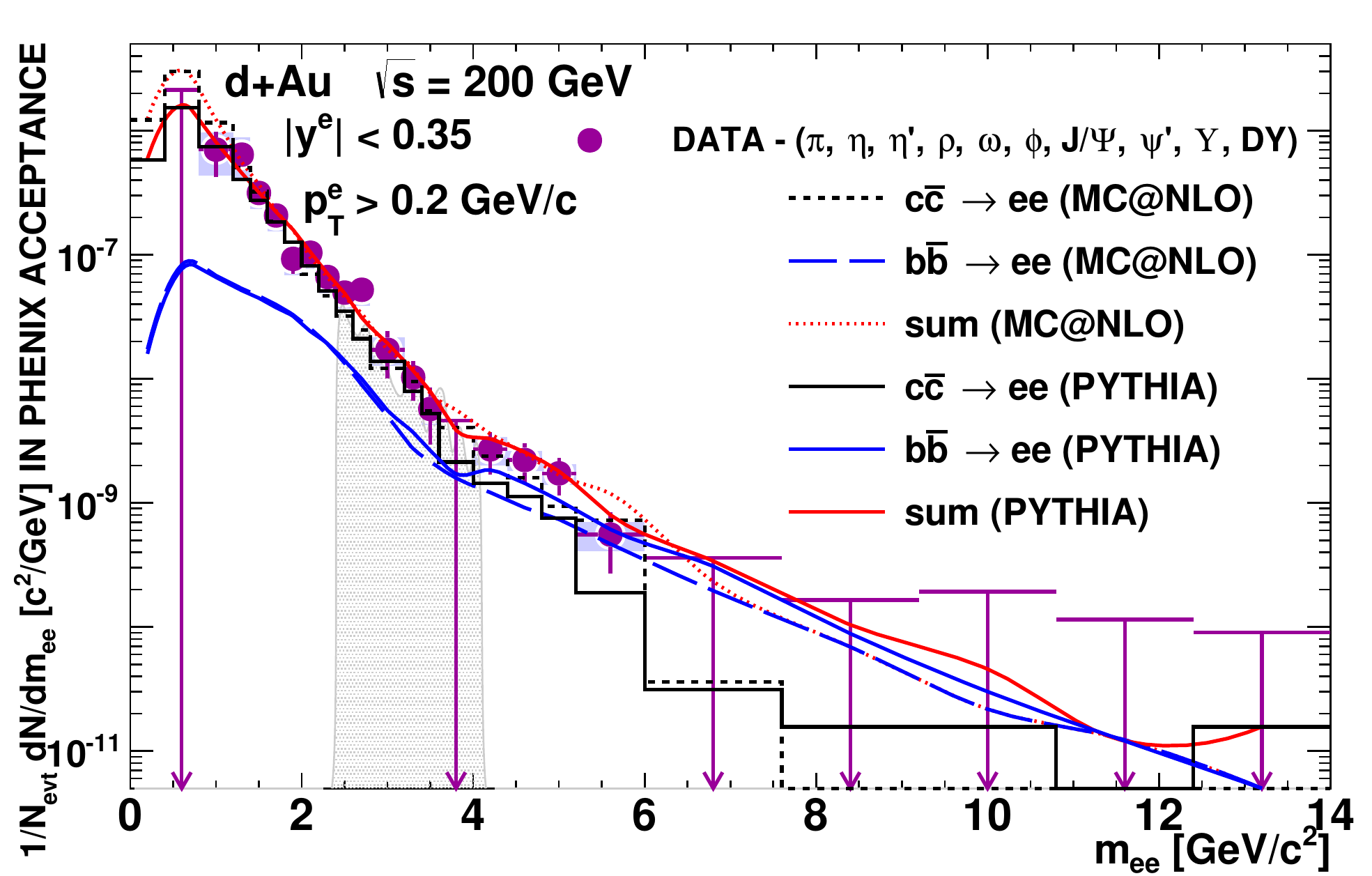}		
\includegraphics[width=1.0\linewidth]{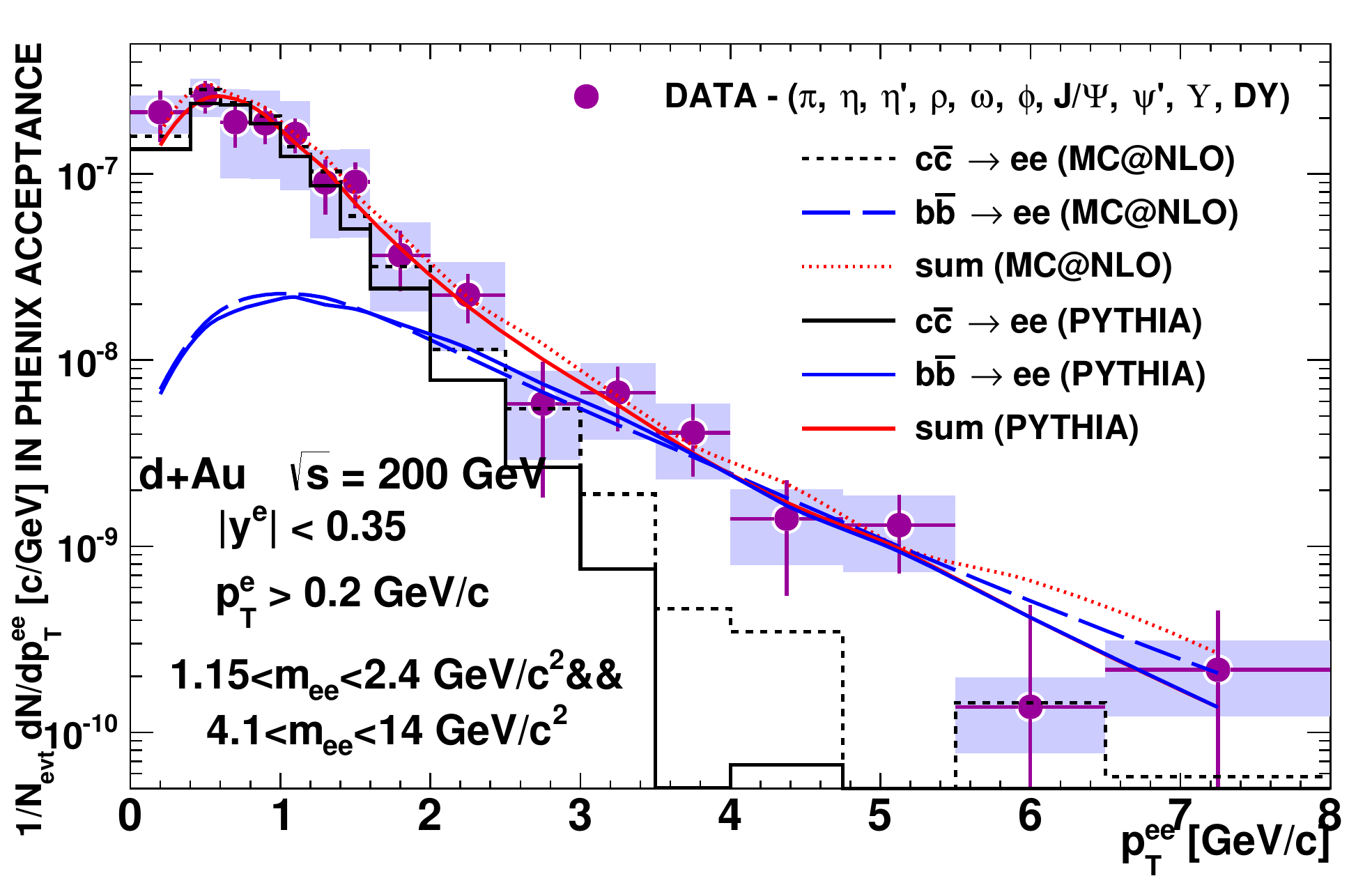}
\caption{\label{Fig:HFprojections}
Top panel compares the mass dependence of \ee pair yield with \pythia 
and \mcnlo calculations. The bottom panel shows the comparison for 
the \pt dependence. The gray panel shown in top panel is not used in 
the fitting and is excluded in the \pt projection.
}
\end{figure}

Figure~\ref{Fig:HFprojections} compares the projections of the \ee 
yield from heavy flavor decays onto the mass and \pt axes to the 
\pythia and \mcnlo calculations. The absolute normalization of 
each calculation was adjusted to the data as discussed below. The 
shape of the measured distributions is well described by both 
simulations. Both projections illustrate the fact that bottom 
production is dominant at high mass or \pt.

\begin{figure}[htb]
\includegraphics[width=1.0\linewidth]{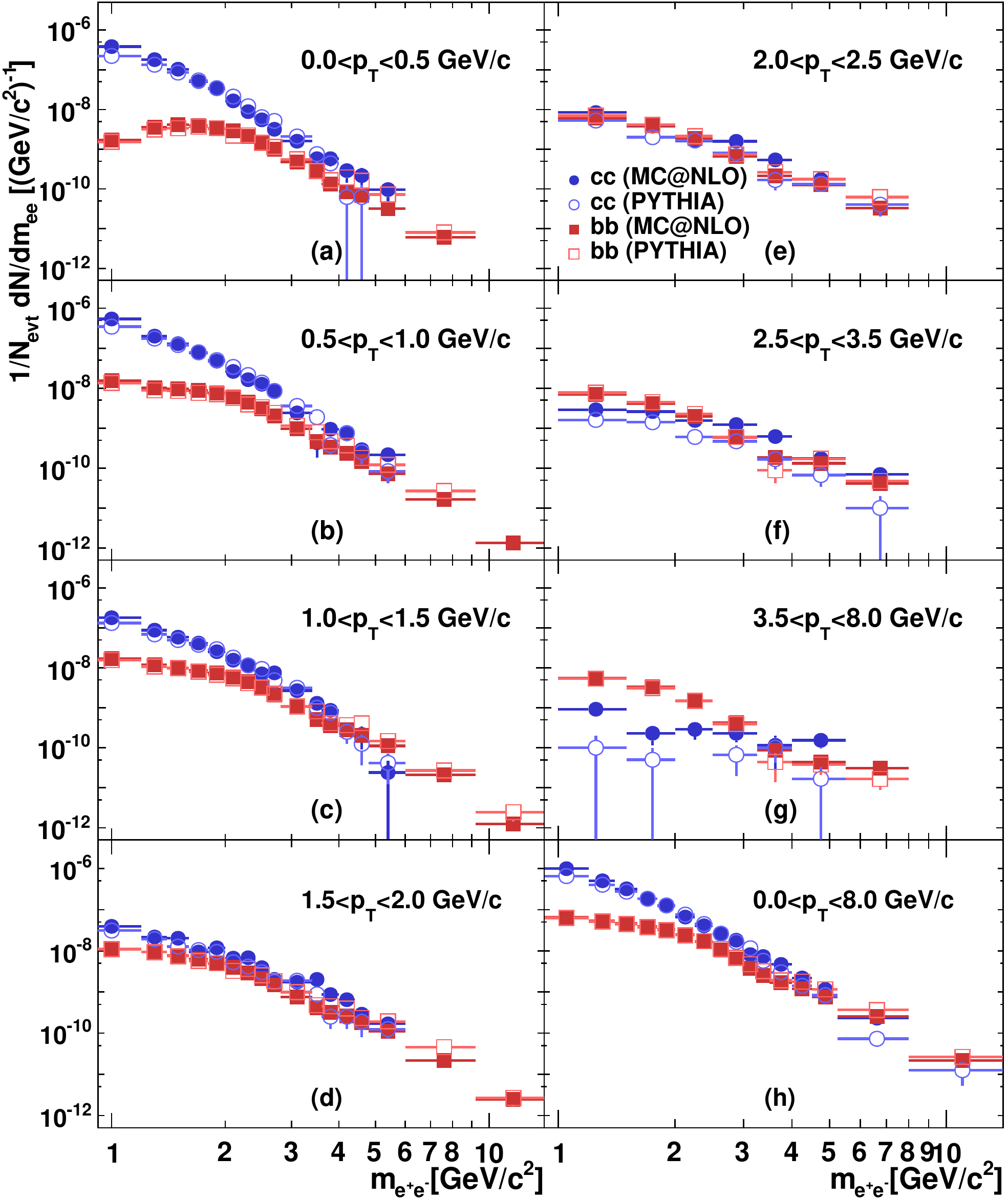}
\caption{\label{Fig:doublediffMC}
Double differential \ee pair yield from semi-leptonic decays of heavy 
flavor as simulated by \pythia and \mcnlo. Shown are mass 
projections in slices of \pt. The \pt intervals are indicated in each 
panel.}
\end{figure}

In the double differential spectra, the separation of \ee pairs from 
charm and bottom decays becomes even more evident. This is 
illustrated in Fig.~\ref{Fig:doublediffMC}. At lower pair momenta, 
charm production dominates the yield below 3 GeV/$c$$^2$ mass. This 
dominance vanishes around \pt = 2 GeV/$c$ and reverses at higher \pt, 
where bottom production dominates. Note that this separation of 
bottom and charm in mass versus \pt is predicted by both generators 
and is thus model independent.

To separate bottom and charm yields quantitatively, we fit the 
distributions shown in Fig.~\ref{Fig:doublediffMC} to the data 
shown in Fig.~\ref{Fig:doublediffdata} with two free parameters, 
$N_{c\bar{c}}$ and $N_{b\bar{b}}$. These, in turn, are used to 
determine the charm and bottom cross sections.

The fits are performed according to
\begin{eqnarray}
\frac {dn_{ee}^{hf}}  {dm dp_T} \Big|_{\rm PHENIX} 
= N_{c\bar{c}} \frac {dn_{ee}^{c\bar{c}} }{dm dp_T} 
+ N_{b\bar{b}} \frac {dn_{ee}^{b\bar{b}} }{dm dp_T}, 
\label{eq:xsec_fit}
\end{eqnarray}
where the left hand side is the measured yield per minimum bias 
triggered event, as shown in Fig.~\ref{Fig:doublediffdata}. The 
$n_{ee}^{c\bar{c}}$ and $n_{ee}^{b\bar{b}}$ are determined either 
using the \pythia simulation or the \mcnlo simulation, where 
the simulation output was normalized to one \cc or \bb pair in 
$4\pi$. The $n_{ee}$ include the branching ratios for both the quark 
and anti-quark to decay semi-leptonically. Furthermore, the simulated 
spectra require that the decay $e^+$ and $e^-$ each have $\pt > 200$ 
MeV/$c$ and that both fall into the PHENIX acceptance and satisfy 
an explicit cut on the pair $m_{T} > 450$~MeV/$c$.  The fits are 
performed in the mass range 
$1.15<m_{e^+e^-}<2.4$~GeV/$c^2$ and $4.1<m_{e^+e^-}<14$~GeV/$c^2$,
for both data and simulations. In this normalization scheme, the fit 
parameters $N_{c\bar{c}}$ and $N_{b\bar{b}}$ are equal to the average 
number \cc pairs and of \bb pairs per inelastic \dA event.

\begin{figure}[tb]
\includegraphics[width=1.0\linewidth]{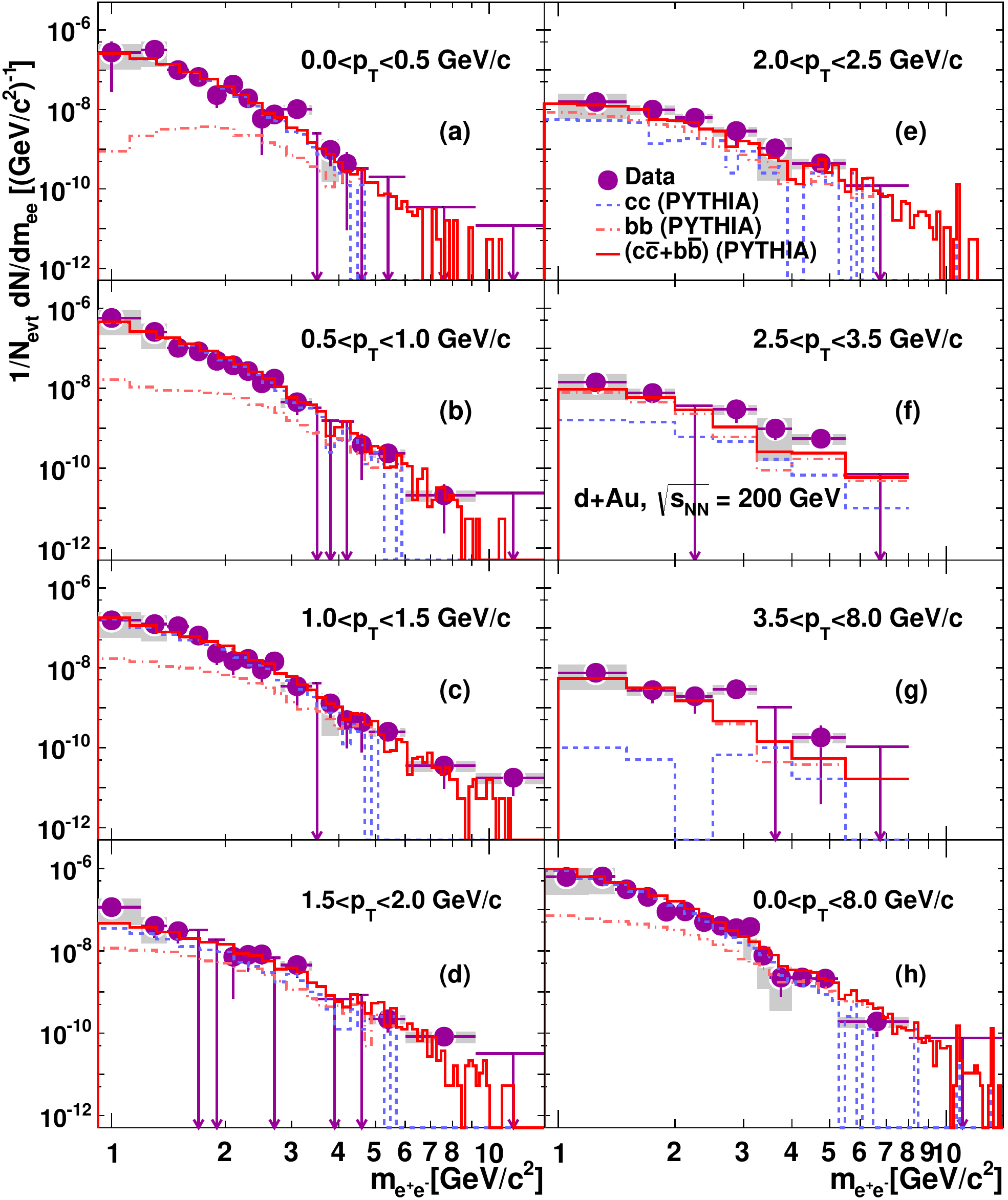}
\caption{\label{Fig:fit_pythia}
Double differential \ee pair yield from heavy flavor decays fitted to 
simulated distributions from \pythia. 
The mass region highlighted by the gray band in 
Fig.~\protect\ref{Fig:HFprojections} is excluded from the fitting.
}
\end{figure}

\begin{figure}[tb]
\includegraphics[width=1.0\linewidth]{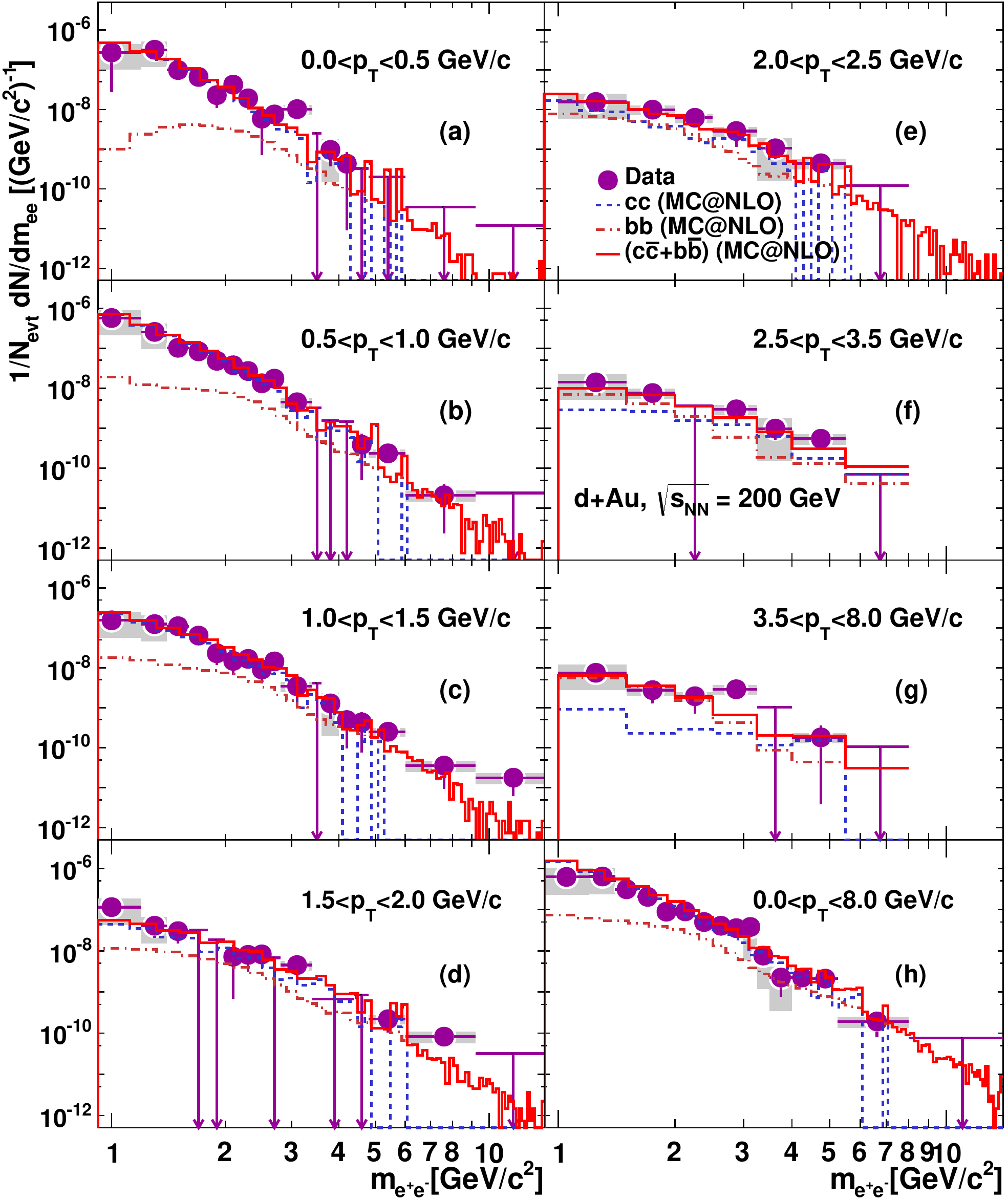}
\caption{\label{Fig:fit_mcnlo}
Double differential \ee pair yield from heavy flavor decays fitted to 
simulated distributions from \mcnlo. 
The mass region highlighted by the gray band in 
Fig.~\protect\ref{Fig:HFprojections} is excluded from the fitting.
}
\end{figure}

The fit results are shown in Fig.~\ref{Fig:fit_pythia} and 
Fig.~\ref{Fig:fit_mcnlo} using the \pythia and \mcnlo 
distributions, respectively. The resulting $\chi^2$ per degree of 
freedom (NDF) is 147/81 for \pythia and 162/81 for \mcnlo. This 
$\chi^2$ is calculated using statistical uncertainty on the data 
points only. If we add the systematic uncertainties in quadrature 
with the systematic uncertainties, the $\chi^2$/NDF is 30/81 and 
34/81 for \pythia and \mcnlo, respectively. These $\chi^2$/NDF 
represent extremes because the statistical uncertainty ignores the 
uncorrelated systematic uncertainty while including the total 
systematic uncertainty incorrectly includes correlated uncertainties. 
Because we do not know the fraction of the correlated and 
uncorrelated systematic uncertainty in the total quoted systematic 
uncertainty, we conservatively assume that it is entirely correlated 
and use the fit results from the corresponding case.

For the \pythia simulation we obtain the fit parameters:
\begin{eqnarray}
N_{c\bar{c}}&=&0.069{\pm}0.006({\rm stat}){\pm}0.021({\rm syst}) \\
N_{b\bar{b}}&=&0.00061{\pm}0.00011({\rm stat}){\pm}0.00019({\rm syst}) 
\end{eqnarray}
and for the \mcnlo 
\begin{eqnarray}
N_{c\bar{c}}&=&0.172{\pm}0.017({\rm stat}){\pm}0.060({\rm syst}) \\
N_{b\bar{b}}&=&0.00060{\pm}0.00014({\rm stat}){\pm}0.00020({\rm syst}) 
\end{eqnarray}
The quoted systematic uncertainties were determined by refitting the 
data points varied up, then down, by one $\sigma_{syst}$.

Additional systematic uncertainties arise from the models themselves. 
In the \mcnlo calculation model uncertainties were evaluated by 
varying the renormalization scale by a factor of 2 up and down; the 
uncertainties are found to be 5\% and 2.5\% for charm and bottom 
respectively. These are quadratically small compared to those arising 
from the data uncertainties. For \pythia no separate evaluation 
of scale-dependence was done.

A second type of model-dependence in the cross section arises from 
the dependence of the pair acceptance on the quark-antiquark 
correlation from the QCD production process, as discussed above. By 
comparing results obtained with the different simulations we can see 
that the model dependence of the bottom cross sections are less than 
2\%. For charm production, on the other hand, the extracted cross 
sections differ by 50\% . The large difference in the model 
dependence of the extracted charm and bottom cross sections results 
from the fact that the bottom mass is much larger and thus the 
fraction of \ee pairs that fall into the PHENIX acceptance is 
dominated by the decay kinematics. For charm production the 
correlation between $c$ and $\bar{c}$ contribute more significantly.

With the fit parameter $N_{b\bar{b}}$ from above, and the acceptance 
relations in Table~\ref{Tab:bb_pairs}, we can determine rapidity 
densities and cross sections for bottom production in \dA collisions. 
The cross section follows as:

\begin{eqnarray}
\sigma^{d\rm{Au}}_{b\bar{b}} = N_{b\bar{b}} \times 
\sigma^{d\rm{Au}}_{\rm inel}  
\end{eqnarray}
We find $1.38 ~\mu$b and $1.36~\mu$b using the $N_{b\bar{b}}$ 
determined using \pythia or \mcnlo, respectively; there is 
essentially no model dependence in the extracted cross sections. 
Consequently, we report the bottom production cross section of:
\begin{eqnarray}
\sigma^{d\rm{Au}}_{b\bar{b}}=1.37{\pm}0.28({\rm stat}){\pm}0.46({\rm syst}) {\rm mb} 
\end{eqnarray}
and a corresponding rapidity density at midrapidity averaged over 
$\Delta y = 1$ of:
\begin{eqnarray}
\frac {d\sigma^{d\rm{Au}}_{b\bar{b}}}{dy}\Big|_{y=0}=0.54{\pm}0.11({\rm stat}){\pm}0.18({\rm syst}) {\rm mb} 
\end{eqnarray}
The average number of binary collisions is $7.6 \pm 0.4$ in inelastic 
\dA events\cite{ppg160}, and the inelastic \pp cross section is $ 
\sigma^{pp}_{\rm inel} = 42 \pm 3 ~m$b. The quoted systematic uncertainty 
on the cross section includes all uncertainties, but is dominated by 
those on the measurement itself.

This is the first measurement of the \bb cross section in $d+$Au 
collisions. One can naively extract a nucleon-nucleon equivalent \bb 
cross section, and find it to be 
$\sigma^{NN}_{bb}=3.4{\pm}0.8({\rm stat}){\pm}1.1({\rm syst})\ \mu$b. 
This value is consistent with the other \bb cross section values as 
reported by other measurements, and a comparison is shown in the 
Table.~\ref{Tab:sigma_bb}.

\begin{table}
\caption{Compilation of the published \bb cross sections.}
    \begin{ruledtabular} \begin{tabular}{cccc}
& $\sigma_{bb}$($\mu \rm b$)        & Reference & \\
      \hline
& 3.4$\pm$0.8 (stat)$\pm$1.1 (syst)  & This work & \\
& 3.2$^{+1.2}_{-1.1}$ (stat)$^{+1.4}_{-1.3}$ (syst)& \cite{ppg094} & \\
& 3.9 $\pm$ 2.5 (stat)$^{+3}_{-2}$ (syst) & \cite{ppg085} & \\
& 4.0 $\pm $0.5 (stat) $\pm$ 1.1 (syst)   & \cite{STARb} & \\
    \end{tabular} \end{ruledtabular}
    \label{Tab:sigma_bb}
\end{table}

Cold-nuclear-matter effects have been measured for heavy flavor in 
\dA \cite{ppg131, PPG130, ppg151, ppg125}. In some cases, the effects 
are small enough to be within the quoted uncertainties of the 
measurement presented here. In others, they occur at forward or 
backward rapidity where the effects will not be observed by these 
data at midrapidity.

The determination of the charm cross section is less reliable due to 
the large model dependence. Using the \pythia calculation we 
find 
$\sigma^{pp}_{c\bar{c}}=385{\pm}34({\rm stat}){\pm}119({\rm syst})\ \mu$b 
and for the \mcnlo calculation we find 
$\sigma^{pp}_{c\bar{c}}=958{\pm}96({\rm stat}){\pm}335({\rm syst})\ \mu$b. 
We conclude that the large model dependence does not allow an accurate 
determination of the charm cross section from our \ee pair 
measurement. As shown in Table~\ref{Tab:cc_ee_pairs}, the model 
dependence of the pair acceptance is already substantial for 
detection of pairs with mass $>$ 1.16 GeV/$c^2$ in one unit of 
rapidity. To test predictions for cold-nuclear-matter effects with 
dilepton data will require comparisons within specific models. 
Calculations should compare the shape of the predicted \ee 
mass and $p_T$ spectra to those presented in 
Fig.~\ref{Fig:doublediffdata} and Fig.~\ref{Fig:HFprojections}.

\section{Summary and Conclusions}

PHENIX recorded a large sample of \ee pairs from \dA collisions at 
\sqsn = 200 GeV in 2008. The \ee pair yield is consistent with the 
expected yield from pseudoscalar and vector meson decays and 
semi-leptonic decays of heavy mesons. The high statistical precision 
of the data allows exploration of both the mass and \pt dependence of 
the \ee yield. Using the double differential information, we can 
clearly isolate the contribution of heavy flavor decays and determine 
the fraction of the yield from \cc and \bb production. We report the 
first measurement of the \bb production cross section in $d+$Au 
collisions.

Our procedure utilizes model predictions of the shape of the double 
differential \ee spectra from \bb and \cc production, with a filter 
requiring that the $e^+$ and $e^-$ fall inside the PHENIX central arm 
acceptance. The two simulations used in this work, \pythia and 
\mcnlo, predict very different correlations between the $q$ and 
$\bar{q}$. In \pythia the \qq correlation is driven by the 
particular implementation of intrinsic $k_T$, while in \mcnlo the \qq 
correlation arises from including NLO terms in the calculation.

For \bb production, the fraction of \ee pairs at midrapidity, and 
therefore also in the PHENIX acceptance, is primarily determined by 
the decay kinematics of the two independent semi-leptonic decays and 
is not sensitive to the substantial model dependence on the \bb 
correlations. For the same reason, the fraction of \ee pairs at 
midrapidity is not sensitive to possible modifications of the momenta 
for $b$ and $\bar{b}$ due to cold-nuclear-matter effects. Determination 
of the \bb cross section thus has little model dependence and the 
measured \ee double differential spectra can be used to reliably
calculate the production cross section, for which we find:
\begin{eqnarray}
\sigma^{d\rm{Au}}_{b\bar{b}}=1.37{\pm}0.28({\rm stat}){\pm}0.46({\rm syst})\ {\rm mb}, 
\end{eqnarray}
A search for cold-nuclear-matter effects will be possible by 
comparing the double differential results reported here with those in 
\pp collisions. The current result should already help to constrain 
models of cold-nuclear-matter effects on heavy quark production.


\section*{ACKNOWLEDGMENTS}


We thank the staff of the Collider-Accelerator and Physics
Departments at Brookhaven National Laboratory and the staff of
the other PHENIX participating institutions for their vital
contributions.  We acknowledge support from the 
Office of Nuclear Physics in the
Office of Science of the Department of Energy, the
National Science Foundation, 
Abilene Christian University Research Council, 
Research Foundation of SUNY, and 
Dean of the College of Arts and Sciences, Vanderbilt University (U.S.A),
Ministry of Education, Culture, Sports, Science, and Technology
and the Japan Society for the Promotion of Science (Japan),
Conselho Nacional de Desenvolvimento Cient\'{\i}fico e
Tecnol{\'o}gico and Funda\c c{\~a}o de Amparo {\`a} Pesquisa do
Estado de S{\~a}o Paulo (Brazil),
Natural Science Foundation of China (P.~R.~China),
Ministry of Education, Youth and Sports (Czech Republic),
Centre National de la Recherche Scientifique, Commissariat
{\`a} l'{\'E}nergie Atomique, and Institut National de Physique
Nucl{\'e}aire et de Physique des Particules (France),
Bundesministerium f\"ur Bildung und Forschung, Deutscher
Akademischer Austausch Dienst, and Alexander von Humboldt Stiftung (Germany),
Hungarian National Science Fund, OTKA (Hungary), 
Department of Atomic Energy and Department of Science and Technology (India), 
Israel Science Foundation (Israel), 
National Research Foundation and WCU program of the 
Ministry Education Science and Technology (Korea),
Physics Department, Lahore University of Management Sciences (Pakistan),
Ministry of Education and Science, Russian Academy of Sciences,
Federal Agency of Atomic Energy (Russia),
VR and Wallenberg Foundation (Sweden), 
the U.S. Civilian Research and Development Foundation for the
Independent States of the Former Soviet Union, 
the US-Hungarian Fulbright Foundation for Educational Exchange,
and the US-Israel Binational Science Foundation.



\end{document}